\documentclass[prx,twocolumn,preprintnumbers,amsmath,amssymb]{revtex4}

\usepackage{graphicx}% Include figure files
\usepackage{bm}% bold math
\usepackage{color}% bold math

\begin{document}

%\preprint{APS/123-QED}

\title{Thermalization in simple metals: The role of electron-phonon and phonon-phonon scatterings}
\author{Shota Ono}
\email{shota\_o@gifu-u.ac.jp}
\affiliation{Department of Electrical, Electronic and Computer Engineering, Gifu University, Gifu 501-1193, Japan}

\begin{abstract}
We study the electron and phonon thermalization in simple metals excited by a laser pulse. The thermalization is investigated numerically by solving the Boltzmann transport equation taking into account all the relevant scattering mechanism: the electron-electron, electron-phonon (e-ph), phonon-electron (ph-e), and phonon-phonon (ph-ph) scatterings. In the initial stage of the relaxation, most of the excitation energy is transferred from the electrons to phonons through the e-ph scattering. This creates hot high-frequency phonons due to the ph-e scatterings, followed by an energy redistribution between phonon subsystems through the ph-ph scatterings. This yields an overshoot of the total longitudinal-acoustic phonon energy at a time, across which a crossover occurs from a nonequilibrium state, where the e-ph and ph-e scatterings frequently occur, to a state, where the ph-ph scattering occurs to reach a thermal equilibrium. This picture is quite different from the scenario of the well-known two-temperature model (2TM). The behavior of the relaxation dynamics is compared with those calculated by several models, including the 2TM, the four-temperature model, and nonequilibrium electron or phonon models. The relationship between the relaxation time and the initial distribution function is also discussed.
\end{abstract}

%\pacs{}

\maketitle

%%%%%%%%%%%%%%%%%%%%%%%%%%%%%%%%%%%%
\section{Introduction}
%%%%%%%%%%%%%%%%%%%%%%%%%%%%%%%%%%%%
Notwithstanding the fundamental interest on ultrafast dynamics of elementary excitations in solids excited by femtosecond laser pulses the thermalization even in simple metals is not well understood. The aim of this paper is to develop a theory for the thermalization of metals beyond the well-known two-temperature model (2TM), that is, beyond the quasiequilibrium approximation. 

The thermalization in metals after a pump pulse irradiation is governed by the electron-electron (e-e), electron-phonon (e-ph), phonon-electron (ph-e), and phonon-phonon (ph-ph) scatterings. The energy transfer between the electrons and phonons through the e-ph and ph-e scatterings has been theoretically studied by Kaganov {\it et al} \cite{kaganov}, motivated by the electron transport experiment, where the deviation from the Ohm's law was observed in metals. Given that the e-e and ph-ph scatterings keep the electron and phonon distributions equal to quasiequilibrium distributions, the energy relaxation can be described by the time-evolution of the electron and phonon temperatures. Based on this picture, Anisimov {\it et al.} have first applied the 2TM to study the energy relaxation of the photoexcited systems \cite{anisimov}. Allen has revealed the relationship between the e-ph coupling function used in the 2TM and the Eliashberg function used in the strong coupling theory of superconductivity \cite{allen}, which has provided how to interpret time-resolved experiments \cite{brorson,perfetti,gadermaier,ono2014,tian2016,sterzi2016} and has been a basis for studying the thermalization of condensed matters theoretically \cite{lin2008,lundgren,brown2016}.

However, it has been questionable whether the assumption behind the 2TM is really valid or not. In fact, numerical studies based on the Boltzmann transport equation (BTE) considering the e-e and e-ph scatterings have revealed the breakdown of the 2TM \cite{groeneveld,rethfeld2002,kabanov2008,ishida2011,mueller,baranov,kemper2017}. Furthermore, recent studies have pointed out the importance of the ph-ph scatterings for better understanding the thermalization in metals \cite{waldecker,ono2017,maldonado,rethfeld2017}, semiconductors \cite{sadasivam}, and Dirac-semimetals \cite{ishida2016}. A picture for the thermalization in solids should now be reconsidered without using the quasiequilibrium approximation.

In this paper, we investigate the electron and phonon thermalization in simple metals by solving the BTE taking into account the e-e, e-ph, ph-e, and ph-ph scatterings. Through the e-ph scatterings, most of the electron energy is transferred into the longitudinal acoustic (LA) phonons in the initial stage of the relaxation. Simultaneously, the LA phonon decays into the transverse acoustic (TA) phonons via the ph-ph scatterings. This yields an overshoot of the total LA phonon energy at a time. Such an overshoot is an indicator for a crossover from a nonequilibrium state, where the e-ph and ph-e scatterings frequently occur, to a state, where the ph-ph scattering occurs to reach a thermal equilibrium. Throughout the energy relaxation, the effect of the e-e scattering can be negligible. The thermalization scenario demonstrated is quite different from that of the 2TM \cite{allen}. A comparative study using several models shows that the energy relaxation of quasiequilibrium states is faster than that of nonequilibrium states. This implies that an application of the 2TM to time-resolved experiments would lead to an underestimation of the e-ph coupling constant of metals, which is consistent with the results in Ref.~\cite{waldecker}. It is also shown that the relaxation time strongly depends on the initial electron and phonon distribution functions as well as their initial energies. 

The paper is organized as follows. In Sec.~\ref{sec:hamiltonian}, the e-e, e-ph, and ph-ph interaction Hamiltonians are formulated. The matrix elements derived are used for describing the multi-particle scattering events. In Sec.~\ref{sec:models}, the BTE for the electron and phonon is derived. For later use, several models for the energy relaxation of solids are provided. In Sec.~\ref{sec:thermalization}, by investigating the time-evolution of the excess energy, the energy transfer rate, and the distribution functions for the electron and phonon, a picture for the thermalization is presented. In Sec.~\ref{sec:noneq}, the behavior of relaxation dynamics in the nonequilibrium electron-phonon model is compared with those in other models. The relationship between the relaxation time and the initial distribution function is also discussed. The paper is summarized in Sec.~\ref{sec:summary}.

%%%%%%%%%%%%%%%%%%%%%%%%%%%%%%%%%%%%
\section{Theory}
\label{sec:theory}
%%%%%%%%%%%%%%%%%%%%%%%%%%%%%%%%%%%%
We consider the nonequilibrium electron-phonon dynamics in simple metals excited by a laser pulse. The low energy excitation of the electron and the phonon is described by the jellium model and the continuum elasticity model, respectively. The ultrafast dynamics of those elementary particles is regulated by the BTE with the e-e, e-ph, ph-e, and ph-ph collision integrals. Thus, we first formulate the interaction Hamiltonian necessary to describe the scattering processes. 

%%%%%%%%%%%%%%%%%%%%%%%%%%%%%%%%%%%%
\subsection{Hamiltonian}
\label{sec:hamiltonian}
%%%%%%%%%%%%%%%%%%%%%%%%%%%%%%%%%%%%
The total Hamiltonian is written as
\begin{eqnarray}
{\cal H}&=& {\cal H}_{\rm e}  + {\cal H}_{\rm p}  + {\cal H}_{\rm ep},
\end{eqnarray}
where ${\cal H}_{\rm e}$, ${\cal H}_{\rm p}$, and ${\cal H}_{\rm ep}$ denote the electron, the phonon, and the e-ph interaction Hamiltonian, respectively. The former two Hamiltonians include the e-e and ph-ph interaction part, respectively. The expression for each Hamiltonian is given below.
%%%%%%%%%%%%%%%%%%%%%%%%%%%%%%%%%%%%
\subsubsection{Electrons}
%%%%%%%%%%%%%%%%%%%%%%%%%%%%%%%%%%%%
In a simple metal, the valence electron can be treated as a free electron in the zeroth-order approximation \cite{ashcroft_mermin}. Since the electrons interact with each other via the Coulomb interaction forces, the electron Hamiltonian is written as \cite{fetter}
\begin{eqnarray}
 {\cal H}_{\rm e} &=& \sum_\sigma \int d\bm{r} \psi_{\sigma}^\dagger (\bm{r}) 
 \left( -\frac{\hbar^2}{2m} \nabla^2 \right)\psi_{\sigma} (\bm{r})
 \nonumber\\
 &+&
 \frac{1}{2} \sum_{\sigma,\sigma'}
  \int d\bm{r} \int d\bm{r}' 
 \psi_{\sigma}^\dagger (\bm{r}) \psi_{\sigma'}^\dagger (\bm{r}') 
V(\vert \bm{x} \vert )
 \psi_{\sigma'} (\bm{r}') \psi_{\sigma} (\bm{r}),
 \nonumber\\
 \label{eq:H_el}
\end{eqnarray}
where $\hbar$ is the Planck constant, $m$ is the electron mass. The first and the second terms are the kinetic and the e-e interaction energies, respectively. $\psi_\sigma (\bm{r})$ is the field operator and is expanded by the plane-waves
\begin{eqnarray}
\psi_{\sigma} (\bm{r}) = \sum_{\bm{k}} a_{\bm{k}\sigma} 
\frac{e^{i\bm{k}\cdot \bm{r}}}{\sqrt{\Omega}}
\end{eqnarray}
with the electron position $\bm{r}$, the wavevector $\bm{k}$, the spin $\sigma$, and the crystal volume $\Omega$. $a_{\bm{k}\sigma}$ ($a_{\bm{k}\sigma}^{\dagger}$) is the destruction (creation) operator of the electron with $\bm{k}$ and $\sigma$. $V(\vert \bm{x} \vert)$ with $\bm{x} = \bm{r} - \bm{r}'$ is the screened Coulomb potential energy given as
\begin{eqnarray}
 V(\vert \bm{x} \vert ) = \frac{e^2}{4\pi \epsilon_0}
 \frac{e^{- q_{\rm TF} \vert \bm{x}\vert }}{ \vert \bm{x} \vert}
\end{eqnarray}
with the electron charge $e$ and the dielectric constant of vacuum $\epsilon_0$. $q_{\rm TF}$ is
the Thomas-Fermi wavenumber given by
\begin{eqnarray}
q_{\rm TF} a_0 &=& \left( \frac{12}{\pi} \right)^{1/3} \frac{1}{\sqrt{r_s}}
%\nonumber\\
%q_{\rm TF} &=& \sqrt{\frac{3e^2 n_0}{2\varepsilon_0 E_F}}
\label{eq:thomas}
\end{eqnarray}
with the dimensionless Wigner-Seitz radius $r_s$ and the Bohr radius $a_0 = 4\pi \varepsilon_0 \hbar^2 / (m e^2)$. Using the Fourier transformation, Eq.~(\ref{eq:H_el}) is expressed as
\begin{eqnarray}
 {\cal H}_{\rm e} &=& \sum_{\bm{k}\sigma} \varepsilon_{\bm{k}}
 a_{\bm{k}\sigma}^{\dagger} a_{\bm{k}\sigma}
 \nonumber\\
 &+&
 \frac{1}{2}\sum_{\bm{k}_1\sigma_1} \sum_{\bm{k}_2\sigma_2} \sum_{\bm{q}}
 \tilde{V}(\bm{q}) 
 a_{\bm{k}_1+\bm{q} \sigma_1}^{\dagger} a_{\bm{k}_2- \bm{q} \sigma_2}^{\dagger}
 a_{\bm{k}_2\sigma_2} a_{\bm{k}_1\sigma_1},
 \nonumber\\
 \label{eq:eHamiltonian}
\end{eqnarray}
where $\varepsilon_{\bm{k}} = \hbar^2 k^2/(2m)$ is the free electron energy and $\tilde{V}(\bm{q})$ is the Fourier component of the screened Coulomb potential 
\begin{eqnarray}
 \tilde{V}(\bm{q}) 
= \frac{1}{\Omega} \frac{e^2}{\epsilon_0 (q^2 +q_{\rm TF}^2)},
\label{eq:scp}
\end{eqnarray}
which is independent of the electron spin. 

%%%%%%%%%%%%%%%%%%%%%%%%%%%%%%%%%%%%
\subsubsection{Phonons}
%%%%%%%%%%%%%%%%%%%%%%%%%%%%%%%%%%%%
The lattice dynamics in an atomistic system can be considered as an elastic wave propagation in a continuum medium within a long wavelength limit \cite{ashcroft_mermin,maradudin}. The Hamiltonian for the later is given by 
\begin{eqnarray}
{\cal H}_{\rm p}
&=& 
\frac{1}{2\rho_{\rm i}} \sum_i \int p_{i}^{*} (\bm{r}) p_i (\bm{r}) d\bm{r} \nonumber\\
&+& 
\frac{1}{2!} \sum_{ijkl}\int d\bm{r} C_{ijkl} 
\eta_{ij}(\bm{r}) \eta_{kl}(\bm{r})
\nonumber\\
&+& \frac{1}{3!} \sum_{ijklmn}
\int d\bm{r} C_{ijklmn} 
\eta_{ij}(\bm{r}) \eta_{kl}(\bm{r}) \eta_{mn}(\bm{r}),
\label{eq:pHamiltonian_real}
\end{eqnarray}
where $C_{ijkl}$ ($C_{ijklmn}$) is the forth-rank (six-rank) tensor and serves as the second-order (third-order) elastic constants with $i,j,k,l,m,n=1, 2, 3$. $\rho_{\rm i}$ is the ion mass density given by $\rho_{\rm i} = M_{\rm i} N_{\rm i}/\Omega$ with the ion mass $M_{\rm i}$ and the number of the ions $N_{\rm i}$ in a volume $\Omega$. The strain tensor is defined as \cite{ziman,landau}
\begin{eqnarray}
\eta_{ij}(\bm{r}) = \frac{1}{2} \left( 
\frac{\partial u_i}{\partial x_j} + \frac{\partial u_j}{\partial x_i}
+
\sum_k \frac{\partial u_k}{\partial x_i} \frac{\partial u_k}{\partial x_j}
\right).
\label{eq:strain}
\end{eqnarray}
Within the isotropic approximation, the forth-rank tensor is given by 
\begin{eqnarray}
C_{ijkl} = 
\lambda_{\rm L} \delta_{ij} \delta_{kl}
+ \mu_{\rm L} (\delta_{ik} \delta_{jl} + \delta_{il} \delta_{jk}),
\label{eq:forth-rank}
\end{eqnarray}
where $\lambda_{\rm L}$ and $\mu_{\rm L}$ are Lam\'{e} constants, while the six-rank tensor is given as
\begin{eqnarray}
& & C_{ijklmn} \nonumber\\
&=& E_1 \delta_{ij} \delta_{kl} \delta_{mn} \nonumber\\
&+& E_2 \Big[ 
\delta_{ij} ( \delta_{km} \delta_{ln} + \delta_{kn} \delta_{lm})
+ \delta_{kl} ( \delta_{im} \delta_{jn} + \delta_{in} \delta_{jm})
\nonumber\\
&+& \delta_{mn} ( \delta_{ik} \delta_{jl} + \delta_{il} \delta_{jk})
\Big]
\nonumber\\
&+& E_3 \Big[
\delta_{ik} ( \delta_{jm} \delta_{ln} + \delta_{jn} \delta_{lm})
+ \delta_{il} ( \delta_{jm} \delta_{kn} + \delta_{jn} \delta_{km})
\nonumber\\
&+& \delta_{im} ( \delta_{jk} \delta_{ln} + \delta_{jl} \delta_{kn})
+ \delta_{in} ( \delta_{jk} \delta_{lm} + \delta_{jl} \delta_{km})
\Big],
%\nonumber\\
%&+& 
%\lambda ( 
%\delta_{ij} \delta_{km} \delta_{ln} + \delta_{im} \delta_{jn} \delta_{kl} + \delta_{ik} \delta_{jl} \delta_{mn})
%\nonumber\\
%&+& 
%\mu( 
%\delta_{ik} \delta_{jm} \delta_{ln} + \delta_{ik} \delta_{jn} \delta_{lm} + \delta_{il} \delta_{jn} \delta_{km}
%\nonumber\\
%&+& 
%\delta_{im} \delta_{jk} \delta_{ln}+ \delta_{im} \delta_{jl} \delta_{kn}+ \delta_{in} \delta_{jl} \delta_{km}
%)
\label{eq:six-rank}
\end{eqnarray}
where $E_1$, $E_2$, and $E_3$ are the third-order elastic constants. The displacement vector is written as 
\begin{eqnarray}
 u_i (\bm{r}) &=& \sum_{\bm{Q},\gamma} \sqrt{\frac{\hbar}{2\rho_{\rm i }\Omega \omega_{\gamma}(\bm{Q})}}
 \left( b_{\bm{Q}\gamma} + b_{ -\bm{Q} \gamma}^{\dagger} \right) 
 \nonumber\\
 &\times&
 \epsilon_i (\bm{Q}, \gamma) e^{i\bm{Q}\cdot \bm{r}},
 \label{eq:displacement}
\end{eqnarray}
where $b_{\bm{Q}\gamma }$ and $b_{\bm{Q}\gamma }^{\dagger}$ are the destruction and creation operators for the phonon with the wavevector $\bm{Q}=(Q_1,Q_2,Q_3)$ and the polarization $\gamma=$LA, TA1, and TA2. $\epsilon_i(\bm{Q}, \gamma)$ is the $i$-th component of the polarization vector $\bm{\epsilon}(\bm{Q}, \gamma)$. The momentum operator is given by
\begin{eqnarray}
 p_i (\bm{r}) &=& -i
 \sum_{\bm{Q},\gamma } \sqrt{\frac{\hbar\rho_{\rm i} \omega_{\gamma}(\bm{Q}) }{2 \Omega }}
 \left( b_{\bm{Q}\gamma } - b_{-\bm{Q} \gamma}^{\dagger} \right) 
 \nonumber\\
 &\times&
 \epsilon_i (\bm{Q},\gamma) e^{i\bm{Q}\cdot \bm{r}}.
\end{eqnarray}
Using the orthonormality of the polarization vectors for a given $\bm{Q}$,
\begin{eqnarray}
\bm{\epsilon}^*(\bm{Q},\gamma') \cdot \bm{\epsilon}(\bm{Q},\gamma) = \delta_{\gamma,\gamma'},
\end{eqnarray}
and the elastic wave equation
\begin{eqnarray}
 \rho_{\rm i} \omega_{\gamma}^2(\bm{Q}) \epsilon_i(\bm{Q},\gamma)
 = \sum_{k} \left( \sum_{jl} C_{ijkl} Q_j Q_l \right) \epsilon_k (\bm{Q},\gamma),
\end{eqnarray}
the unperturbed phonon Hamiltonian, that is, the sum of the first and second terms in Eq.~(\ref{eq:pHamiltonian_real}), is given by
\begin{eqnarray}
 {\cal H}_{p}^{0} = \sum_{\bm{Q}\gamma} 
 \hbar \omega_{\gamma}(\bm{Q}) 
 \left( b_{\bm{Q} \gamma}^{\dagger} b_{\bm{Q} \gamma} + \frac{1}{2}\right).
 \end{eqnarray}
The frequencies of the three phonon branches are given by 
\begin{eqnarray}
\omega_{{\rm LA}}(\bm{Q}) &=& 
\sqrt{\frac{\lambda_{\rm L} + 2\mu_{\rm L}}{\rho_{\rm i}}} \vert \bm{Q} \vert 
\equiv v_{\rm LA} \vert \bm{Q} \vert,
\nonumber\\
\omega_{{\rm TA1}}(\bm{Q}) 
&=& \omega_{{\rm TA2}}(\bm{Q}) 
= \sqrt{\frac{\mu_{\rm L}}{\rho_{\rm i}}} \vert \bm{Q} \vert
 \equiv v_{\rm TA} \vert \bm{Q} \vert,
\end{eqnarray}
where $v_{\rm LA}$ and $v_{\rm TA}$ are the phonon velocities. The Debye frequency $\Omega_{\gamma, {\rm D}}$ for the polarization $\gamma$ is determined by the normalization condition $N_{\rm i} = \int_{0}^{\Omega_{\gamma, {\rm D}}} {\cal D}_\gamma (\omega) d\omega$, where ${\cal D}_\gamma (\omega)$ is the density-of-states (DOS) for phonons, resulting in $\Omega_{\gamma, {\rm D}}=v_\gamma (6\pi^2 N_{\rm i}/\Omega)^{1/3}$.

The time-evolution of the phonon population is governed by the three-phonon process. Substituting Eq.~(\ref{eq:displacement}) into Eq.~(\ref{eq:pHamiltonian_real}) and defining
\begin{eqnarray}
 \tilde{\cal C}_{ijklmn}
&=& C_{ijlkmn}
\nonumber\\
&+& 
\lambda_{\rm L} ( 
\delta_{ij} \delta_{km} \delta_{ln} 
+ \delta_{im} \delta_{jn} \delta_{kl} 
+ \delta_{ik} \delta_{jl} \delta_{mn})
\nonumber\\
&+& 
\mu_{\rm L} ( 
\delta_{ik} \delta_{jm} \delta_{ln}
+ \delta_{ik} \delta_{jn} \delta_{lm}
+ \delta_{il} \delta_{jn} \delta_{km}
\nonumber\\
&+& 
\delta_{im} \delta_{jk} \delta_{ln}
+ \delta_{im} \delta_{jl} \delta_{kn}
+ \delta_{in} \delta_{jl} \delta_{km}
),
\label{eq:six-rank2}
\end{eqnarray}
one obtains the perturbed Hamiltonian 
\begin{eqnarray}
 {\cal H}'_{p} &=& 
 \frac{1}{6}\sum_{\bm{Q},\bm{Q}',\bm{Q}''} \sum_{\gamma,\gamma',\gamma'}
 A_{\bm{Q},\bm{Q}',\bm{Q}''}^{\gamma,\gamma',\gamma''}
 \delta_{\bm{Q}+\bm{Q}' + \bm{Q}'',0}
  \nonumber\\
  &\times&
 B_{\bm{Q}\gamma}
 B_{\bm{Q}'\gamma'}
 B_{\bm{Q}''\gamma''},
 \label{eq:pHamiltonian}
\end{eqnarray}
where $B_{\bm{Q}\gamma} = b_{\bm{Q}\gamma} +  b_{-\bm{Q}\gamma}^{\dagger}$ and the three-phonon matrix elements $A_{\bm{Q},\bm{Q}',\bm{Q}''}^{\gamma,\gamma',\gamma'}$ that is explicitly given as \cite{tamura}
%\begin{eqnarray}
% A_{\bm{Q},\bm{Q}',\bm{Q}''}^{\gamma,\gamma',\gamma''}
% &=&
 % \frac{-i}{\sqrt{\Omega}}
 % \left( \frac{\hbar}{2\rho}\right)^{3/2}
 %\sum_{ijklmn}
 %\tilde{\cal C}_{ijklmn}
 %\nonumber\\
% &\times&
 % \frac{
 % e_{i}(\bm{Q}\gamma) 
 % e_{k}(\bm{Q}'\gamma') 
 % e_{m}(\bm{Q}''\gamma'') 
 % Q_j Q'_{l} Q''_{n}}
 % {\sqrt{\omega_\gamma (\bm{Q})  \omega_{\gamma'} (\bm{Q}')  \omega_{\gamma''} (\bm{Q}'')  }}
  % \label{eq:3phonon_matrix}
%\end{eqnarray}
%By inserting Eq.~(\ref{eq:six-rank2}) into Eq.~(\ref{eq:3phonon_matrix}), one obtains \cite{tamura}
\begin{eqnarray}
 A_{\bm{Q},\bm{Q}',\bm{Q}''}^{\gamma,\gamma',\gamma''}
 &=&
  \frac{-i}{\sqrt{\Omega}}
  \left( \frac{\hbar}{2\rho_{\rm i}}\right)^{3/2}
   \frac{
   A_{\rm ph}
  }
  {\sqrt{\omega_\gamma (\bm{Q})  \omega_{\gamma'} (\bm{Q}')  \omega_{\gamma''} (\bm{Q}'')  }}
  \nonumber\\
 \end{eqnarray}
with 
\begin{widetext}
\begin{eqnarray}
 A_{\rm ph} &=& 
 E_1 (\bm{e} \cdot \bm{Q})  (\bm{e}' \cdot \bm{Q}') (\bm{e}'' \cdot \bm{Q}'')
 \nonumber\\
&+&  E_2 (\bm{e} \cdot \bm{Q}) \left[
 (\bm{e}' \cdot \bm{e}'') (\bm{Q}' \cdot \bm{Q}'')
+ (\bm{e}' \cdot \bm{Q}'') (\bm{e}'' \cdot \bm{Q}') \right]
+  E_2 (\bm{e}' \cdot \bm{Q}')  \left[
(\bm{e} \cdot \bm{e}'') (\bm{Q} \cdot \bm{Q}'')
+ (\bm{e} \cdot \bm{Q}'') (\bm{e}'' \cdot \bm{Q}) \right]
 \nonumber\\
&+&  E_2 (\bm{e}'' \cdot \bm{Q}'')  \left[
(\bm{e} \cdot \bm{e}') (\bm{Q} \cdot \bm{Q}')
+ (\bm{e} \cdot \bm{Q}') (\bm{e}' \cdot \bm{Q}) \right]
\nonumber\\
&+& E_3 (\bm{e} \cdot \bm{e}') \left[
 (\bm{e}'' \cdot \bm{Q}) (\bm{Q}' \cdot \bm{Q}'')
+ (\bm{Q} \cdot \bm{Q}'') (\bm{e}'' \cdot \bm{Q}') \right]
+  E_3 (\bm{e} \cdot \bm{Q}')  \left[
(\bm{e}'' \cdot \bm{Q}) (\bm{e}' \cdot \bm{Q}'')
+ (\bm{Q} \cdot \bm{Q}'') (\bm{e}' \cdot \bm{e}'') \right]
\nonumber\\
&+& E_3 (\bm{e} \cdot \bm{e}'') \left[
 (\bm{e}' \cdot \bm{Q}) (\bm{Q}' \cdot \bm{Q}'')
+ (\bm{Q} \cdot \bm{Q}') (\bm{e}' \cdot \bm{Q}'') \right]
+  E_3 (\bm{e} \cdot \bm{Q}'')  \left[
(\bm{e}' \cdot \bm{Q}) (\bm{e}'' \cdot \bm{Q}')
+ (\bm{Q} \cdot \bm{Q}') (\bm{e}' \cdot \bm{e}'') \right]
\nonumber\\
%--------------------------------------------%
&+& 
\lambda_{\rm L} \left[ 
(\bm{e} \cdot \bm{Q}) (\bm{e}' \cdot \bm{e}'') (\bm{Q}' \cdot \bm{Q}'')
+ (\bm{e} \cdot \bm{e}'') (\bm{Q} \cdot \bm{Q}'') (\bm{e}' \cdot \bm{Q}')
+ (\bm{e} \cdot \bm{e}') (\bm{Q} \cdot \bm{Q}') (\bm{e}'' \cdot \bm{Q}'')
\right]
%--------------------------------------------%
\nonumber\\
%--------------------------------------------%
&+& 
\mu_{\rm L} \big[ 
(\bm{e} \cdot \bm{e}') (\bm{Q} \cdot \bm{e}'') (\bm{Q}' \cdot \bm{Q}'')
+ (\bm{e} \cdot \bm{e}') (\bm{Q} \cdot \bm{Q}'') (\bm{Q}' \cdot \bm{e}'')
+ (\bm{e} \cdot \bm{Q}') (\bm{Q} \cdot \bm{Q}'') (\bm{e}' \cdot \bm{e}'')
%--------------------------------------------%
\nonumber\\
%--------------------------------------------%
&+& 
(\bm{e} \cdot \bm{e}'') (\bm{Q} \cdot \bm{e}') (\bm{Q}' \cdot \bm{Q}'')
+ (\bm{e} \cdot \bm{e}'') (\bm{Q} \cdot \bm{Q}') (\bm{e}' \cdot \bm{Q}'')
+ (\bm{e} \cdot \bm{Q}'') (\bm{Q} \cdot \bm{Q}') (\bm{e}' \cdot \bm{e}'')
\big]
%--------------------------------------------%
\label{eq:A_ph}
\end{eqnarray}
\end{widetext}
with the abbreviation $\bm{e}= \bm{e}(\bm{Q}\gamma)$, $\bm{e}'=\bm{e}(\bm{Q}'\gamma')$, and $\bm{e}''=\bm{e}(\bm{Q}''\gamma'')$. Note that $A_{\rm ph}$ is symmetric under the exchanges $(\bm{e},\bm{Q}) \leftrightarrow (\bm{e}',\bm{Q}')$, $(\bm{e}',\bm{Q}') \leftrightarrow (\bm{e}'',\bm{Q}'')$, and $(\bm{e},\bm{Q}) \leftrightarrow (\bm{e}'',\bm{Q}'')$.

%%%%%%%%%%%%%%%%%%%%%%%%%%%%%%%%%%%%
\subsubsection{Electron-Phonon Interaction}
%%%%%%%%%%%%%%%%%%%%%%%%%%%%%%%%%%%%
The electron-lattice interaction Hamiltonian is expanded into two terms: the static and dynamical lattice potential. The former and the latter contribute to the Bloch electron formation and the energy exchange during the relaxation. The leading term in the latter is the deformation potential interaction \cite{smith}. The e-ph interaction Hamiltonian is given by 
\begin{eqnarray}
{\cal H}_{\rm ep} &=& \sum_{\sigma} \int d\bm{r} \psi_{\sigma}^{\dagger} (\bm{r})
\left( D_0 \bm{\nabla} \cdot \bm{u} \right)
\psi_{\sigma} (\bm{r})
\nonumber\\
&=& 
\sum_{\bm{k},\sigma} \sum_{\bm{Q},\gamma} g(\bm{Q},\gamma)
a_{\bm{k}+\bm{Q}\sigma}^{\dagger}
a_{\bm{k}\sigma} B_{\bm{Q}\gamma}
\label{eq:ephHamiltonian}
\end{eqnarray}
where $D_0$ is the deformation potential, which describes the interaction between the electron and the LA phonon through the local density fluctuations of continuum medium. Within the free-electron approximation, $ D_0 = 2 \varepsilon_F /3$ with $\varepsilon_F$ being the Fermi energy. Then, one obtains
\begin{eqnarray}
\vert g (\bm{Q},\gamma) \vert^2
= D_{0}^{2}
\frac{\hbar \vert \bm{Q}\vert}{2\rho_{\rm i}\Omega v_{\rm LA}} \delta_{\gamma,{\rm LA}}.
\end{eqnarray}
The matrix elements of the electron-TA phonon interaction is finite when one considers the umklapp processes. To treat it phenomenologically, we introduce the polarization-dependent deformation potential $D_\gamma$ and assume
\begin{eqnarray}
\vert g(\bm{Q},\gamma) \vert^2
= D_{\gamma}^{2}
\frac{\hbar \vert \bm{Q}\vert }{2\rho_{\rm i}\Omega v_{\gamma}},
\label{eq:eph_matrix}
\end{eqnarray}
where $D_{\rm LA}= D_0$ and $D_{\rm TA}= (v_{\rm TA}/v_{\rm LA})^\beta D_0$. The parameter $\beta$ is determined to yield a realistic value of the e-ph coupling constant described below. 

%%%%%%%%%%%%%%%%%%%%%%%%%%%%%%%%%%%%
\subsection{Models}
\label{sec:models}
%%%%%%%%%%%%%%%%%%%%%%%%%%%%%%%%%%%%
We will formulate five models depending on the approximation level. In the first model given in Sec.~\ref{sec:NEP}, the presence of the quasiequilibrium states is not assumed, while in the 2TM given in Sec.~\ref{sec:2TM} it is assumed {\it a priori}. 
%%%%%%%%%%%%%%%%%%%%%%%%%%%%%%%%%%%%
\subsubsection{NEP model}
\label{sec:NEP}
%%%%%%%%%%%%%%%%%%%%%%%%%%%%%%%%%%%%
We first present the nonequilibrium electron and phonon (NEP) model, where all the relevant scattering processes are considered. We thus expect that the NEP model would yield a relaxation dynamics quite similar to the ultrafast dynamics observed in time-resolved experiments. The BTE for electron and phonon systems is written as \cite{ziman}
\begin{eqnarray}
 \frac{\partial f_{\bm{k},\sigma}}{\partial t} 
 &=& \left( \frac{\partial f}{\partial t} \right)_{\rm e \mathchar`- e}
 + \left( \frac{\partial f}{\partial t} \right)_{\rm e \mathchar`- ph},
 \label{eq:eBTE}
\\
 \frac{\partial n_{\bm{Q},\gamma}}{\partial t} 
 &=& \left( \frac{\partial n}{\partial t} \right)_{\rm ph \mathchar`- e}
 + \left( \frac{\partial n}{\partial t} \right)_{\rm ph \mathchar`- ph},
  \label{eq:pBTE}
 \end{eqnarray}
where no contribution from the diffusion and external field terms is assumed. $f_{\bm{k},\sigma}$ denotes the distribution function of the electron state with $(\bm{k},\sigma)$. $n_{\bm{Q},\gamma}$ denotes the distribution function of the phonon state with $(\bm{Q},\gamma)$. The first and second terms in Eq.~(\ref{eq:eBTE}) indicate the e-e and e-ph collision integrals, respectively, while the first and second terms in Eq.~(\ref{eq:pBTE}) indicate the ph-e and ph-ph collision integrals, respectively. 

Given no magnetic impurities and weak exchange interaction between the electrons in the system, it is reasonable to assume that the electron distribution is independent of the spin coordinate, that is, 
\begin{eqnarray}
 f_{\bm{k},\uparrow} = f_{\bm{k},\downarrow} \equiv f_{\bm{k}}.
\end{eqnarray}

The transition probability for the scattering events is formulated within the Fermi's golden rule. Using Eqs.~(\ref{eq:eHamiltonian}), (\ref{eq:pHamiltonian}), (\ref{eq:ephHamiltonian}), and (\ref{eq:eph_matrix}), the e-e, e-ph, ph-e, and ph-ph collision integrals are written as 
\begin{widetext}
\begin{eqnarray}
%%%%%%%%%%%%%%%%%%%%%%%%%%%%%%%%
% ELECTRON BTE (k - representation)
%%%%%%%%%%%%%%%%%%%%%%%%%%%%%%%%
  \left( \frac{\partial f}{\partial t} \right)_{\rm e \mathchar`- e}
  &=& \sum_{\bm{k}',\bm{q}} \frac{2\pi}{\hbar} \vert \tilde{V}(\bm{q}) \vert^2 
  \delta (\Delta \varepsilon)
  \left[ 
  - f_{\bm{k}} f_{\bm{k}'} (1- f_{\bm{k} + \bm{q}}) (1- f_{\bm{k}'- \bm{q}})
  + (1- f_{\bm{k}}) (1- f_{\bm{k}'}) f_{\bm{k}+\bm{q}} f_{\bm{k}' - \bm{q}}
  \right],
  \label{eq:coll_e-e}
  \\
%%%%%%%%%%%%%%%%%%%%%%
    \left( \frac{\partial f}{\partial t} \right)_{\rm e \mathchar`- ph}
  &=& \sum_{\bm{Q},\gamma} \frac{2\pi}{\hbar} \vert g(\bm{Q},\gamma) \vert^2 
  \Big\{ 
  - f_{\bm{k}} (1-f_{\bm{k}+\bm{Q}} )
  \left[ n^{(+)}_{\bm{Q}}
  \delta ( \varepsilon_{\bm{k}} -  \varepsilon_{\bm{k}+\bm{Q}} -  \hbar\omega_{\bm{Q}\gamma}) 
  + n_{\bm{Q}}
  \delta ( \varepsilon_{\bm{k}} -  \varepsilon_{\bm{k}+\bm{Q}} +  \hbar\omega_{\bm{Q}\gamma}) 
  \right]
\nonumber\\
 & & + (1- f_{\bm{k}}) f_{\bm{k}+\bm{Q}} 
  \left[ n^{(+)}_{\bm{Q}}
  \delta ( \varepsilon_{\bm{k}} -  \varepsilon_{\bm{k}+\bm{Q}} + \hbar\omega_{\bm{Q}\gamma}) 
  + n_{\bm{Q}}
  \delta ( \varepsilon_{\bm{k}} -  \varepsilon_{\bm{k}+\bm{Q}} -  \hbar\omega_{\bm{Q}\gamma}) 
  \right]
  \Big\}, 
  \label{eq:coll_e-ph}
 \\
 %%%%%%%%%%%%%%%%%%%%%%%%%%%%%%%%
% PHONON BTE (k - representation)
%%%%%%%%%%%%%%%%%%%%%%%%%%%%%%%%
    \left( \frac{\partial n}{\partial t} \right)_{\rm ph \mathchar`- e}
  &=& \sum_{\bm{k}} \frac{4\pi}{\hbar} \vert g(\bm{Q},\gamma) \vert^2 
  f_{\bm{k}} (1-f_{\bm{k}+\bm{Q}} )
  \left[ - n_{\bm{Q},\gamma}
  \delta ( \varepsilon_{\bm{k}} -  \varepsilon_{\bm{k}+\bm{Q}} + \hbar\omega_{\bm{Q}\gamma}) 
  + n^{(+)}_{\bm{Q},\gamma}
  \delta ( \varepsilon_{\bm{k}} -  \varepsilon_{\bm{k}+\bm{Q}} -  \hbar\omega_{\bm{Q}\gamma}) 
  \right],
  \label{eq:coll_ph-e}
  \\
%%%%%%%%%%%%%%%%%%%%%%
    \left( \frac{\partial n}{\partial t} \right)_{\rm ph \mathchar`- ph}
 &=&  \sum_{\bm{Q}' \gamma'} \sum_{\gamma''}
 \frac{2\pi}{\hbar} 
 \left\vert A_{\bm{Q},\bm{Q}',-(\bm{Q}+\bm{Q}')}^{\gamma,\gamma',\gamma''} \right\vert^2
 \nonumber\\
 &\times&
 \Big\{
 \frac{1}{2} 
 \left[ n_{\bm{Q},\gamma}^{(+)} n_{-\bm{Q}',\gamma'} n_{\bm{Q}+\bm{Q}',\gamma''}
 - n_{\bm{Q},\gamma} n_{-\bm{Q}',\gamma'}^{(+)} n_{\bm{Q}+\bm{Q}',\gamma''}^{(+)} \right]
 \delta\left(\hbar\omega_{\bm{Q},\gamma} - \hbar\omega_{-\bm{Q}',\gamma'} -\hbar\omega_{\bm{Q}+\bm{Q}',\gamma''} \right)
 \nonumber\\
 & + &
\left[ n_{\bm{Q},\gamma}^{(+)} n_{\bm{Q}',\gamma'}^{(+)} n_{\bm{Q}+\bm{Q}',\gamma''}
 - n_{\bm{Q},\gamma} n_{\bm{Q}',\gamma'} n_{\bm{Q}+\bm{Q}',\gamma''}^{(+)} \right]
 \delta\left(\hbar\omega_{\bm{Q},\gamma} + \hbar\omega_{\bm{Q}',\gamma'} -\hbar\omega_{\bm{Q}+\bm{Q}',\gamma''} \right)
 \Big\},
 \label{eq:coll_ph-ph}
%%%%%%%%%%%%%%%%%%%%%%
 \end{eqnarray}
 \end{widetext}
where $\Delta \varepsilon = \varepsilon_{\bm{k}} + \varepsilon_{\bm{k}'}  - \varepsilon_{\bm{k}+ \bm{q}} - \varepsilon_{\bm{k}' - \bm{q}}$ and $n_{\bm{Q},\gamma}^{(+)} = n_{\bm{Q},\gamma} + 1$. Eq.~(\ref{eq:coll_e-e}) describes the electron scattering $(\bm{k},\bm{k}') \leftrightarrows (\bm{k}+\bm{q},\bm{k}'-\bm{q})$ governed by the screened Coulomb interaction potential $\tilde{V}(\bm{q})$ in Eq.~(\ref{eq:scp}). Eqs.~(\ref{eq:coll_e-ph}) and (\ref{eq:coll_ph-e}) describes the e-ph and ph-e scatterings, where the electron with $\bm{k}$ is scattered into that with $\bm{k}+\bm{Q}$ by an absorption of the phonon with $\bm{Q}$ or an emission of the phonon with $-\bm{Q}$, and vice versa. The first and second terms in Eq.~(\ref{eq:coll_ph-ph}) denote the phonon anharmonic decay and inelastic scatterings, respectively, with the total wavevector conserved. For the former process, the phonon mode with $(\bm{Q},\gamma)$ decays into two phonons of $(\bm{Q}',\gamma')$ and $(\bm{Q} - \bm{Q}',\gamma'')$, while for the latter one, the two phonon modes with $(\bm{Q},\gamma)$ and $(\bm{Q}',\gamma')$ are merged into a phonon with $(\bm{Q} + \bm{Q}',\gamma'')$. In the derivation of the collision terms, Eqs.~(\ref{eq:coll_e-ph}), (\ref{eq:coll_ph-e}), and (\ref{eq:coll_ph-ph}), the relations $\omega_{\bm{Q}\gamma} = \omega_{-\bm{Q}\gamma}$ and $n_{\bm{Q},\gamma} = n_{-\bm{Q},\gamma}$ arising from the inversion symmetry are used.

Note that, for a simple metal, there is no three-phonon processes in which all three phonons belong to the same polarization branches. This is due to the dispersion effect near the zone boundary \cite{ziman}. In the present study, one can neglect the three-phonon processes, such as LA (TA) $\leftrightarrows$ LA (TA) $+$ LA (TA). Besides, there is no scattering processes, in which one TA phonon creates two LA phonons, and vice versa, due to the energy conservation law \cite{ziman}. 

When $f_{\bm{k}}$ is averaged over the electron states having the energy $\varepsilon$, one obtains the distribution function for the electron state with the energy $\varepsilon$ \cite{kabanov2008,baranov}
\begin{eqnarray}
 f (\varepsilon) = \frac{1}{{\cal N}(\varepsilon)} \sum_{\bm{k}} \delta (\varepsilon - \varepsilon_{\bm{k}}) f_{\bm{k}},
 \label{eq:fepsi}
\end{eqnarray}
where ${\cal N}(\varepsilon) = \Omega (2m)^{3/2} \sqrt{\varepsilon}/(4\pi^2\hbar^3)$ is the electron DOS per spin. Similarly, the phonon distribution function for the phonon states with the frequency $\omega$ is defined as \cite{kabanov2008,baranov}
\begin{eqnarray}
 n_\gamma (\omega) = \frac{1}{{\cal D}_\gamma(\omega)} \sum_{\bm{Q}} \delta (\omega - \omega_{\bm{Q}\gamma}) n_{\bm{Q},\gamma},
  \label{eq:nomega}
\end{eqnarray}
where ${\cal D}_\gamma (\omega) = \Omega \omega^2/(2\pi^2 v_{\gamma}^3) \theta_{\rm H}(\Omega_{ \gamma,{\rm D}} - \omega)$ is the phonon DOS for the polarization $\gamma$. $\theta_{\rm H} (x)$ is the Heaviside step function: $\theta (x) = 1$ for $x\ge 1$ and $\theta (x) = 0$ for $x <1$. By using Eqs.~(\ref{eq:fepsi}) and (\ref{eq:nomega}), the time ($t$)-evolution for the electron and phonon distribution functions are given by, respectively,
\begin{widetext}
\begin{eqnarray}
%%%%%%%%%%%%%%%%%%%%%%%%%%%%%%%%
% ELECTRON BTE (energy representation)
%%%%%%%%%%%%%%%%%%%%%%%%%%%%%%%%
\frac{\partial f (\varepsilon)}{\partial t}
&=& 2\pi \int d\varepsilon' \int d\xi \int d\xi'  C_{\rm e \mathchar`- e}(\varepsilon,\varepsilon',\xi,\xi') 
\delta (\varepsilon + \varepsilon' - \xi - \xi') 
\nonumber\\
&\times&
\left\{ 
- f (\varepsilon) f (\varepsilon' ) [1- f(\xi)] [1- f(\xi' )]
+ [1- f (\varepsilon)] [1- f (\varepsilon' )] f(\xi) f(\xi' )
\right\}
\nonumber\\
%%%%%%%%%%%%%%%%
&+& 2\pi \sum_{\gamma} \int d\xi \int d\omega  C_{\rm e \mathchar`- ph}(\varepsilon, \xi,\omega,\gamma) 
\nonumber\\
&\times& 
\Big(
\delta (\varepsilon  - \xi - \hbar\omega) 
\left\{ 
[f(\xi) - f (\varepsilon)] n_\gamma(\omega) 
- f (\varepsilon) [1- f(\xi )]
\right\}
+ 
\delta (\varepsilon  - \xi + \hbar\omega) 
\left\{ 
[f(\xi) - f (\varepsilon)] n_\gamma(\omega) 
+ f(\xi ) [1- f (\varepsilon)]
\right\}
\Big)
\nonumber\\
\label{eq:edist}
\end{eqnarray}
and 
%%%%%%%%%%%%%%%%%%%%%%%%%%%%%%%%
% PHONON BTE (frequency representation)
%%%%%%%%%%%%%%%%%%%%%%%%%%%%%%%%
\begin{eqnarray}
\frac{\partial n_\gamma (\omega)}{\partial t}
&=& 4\pi \int d\varepsilon \int d\xi  C_{\rm ph \mathchar`- e}(\varepsilon, \xi,\omega,\gamma) 
f (\varepsilon) [1- f(\xi)]
\left\{
- n_\gamma(\omega) \delta (\varepsilon - \xi + \hbar \omega)
+ [n_\gamma(\omega)+1] \delta (\varepsilon - \xi - \hbar \omega)
\right\}
\nonumber\\
%%%%%%%%%%%%%%%%
&+&  
2\pi \sum_{\gamma'} \sum_{\gamma''} \int d\omega' \int d\omega''  
C_{\rm ph \mathchar`- ph}(\omega, \omega', \omega'', \gamma, \gamma', \gamma'') 
\nonumber\\
&\times&
\Big\{
\frac{1}{2} 
\left[
n_{\gamma}^{(+)}(\omega) n_{\gamma'}(\omega') n_{\gamma}''(\omega'') 
- n_{\gamma}(\omega) n_{\gamma'}^{(+)}(\omega') n_{\gamma''}^{(+)}(\omega'') 
\right]
 \delta (\hbar \omega - \hbar \omega' - \hbar \omega'')
\nonumber\\
&+&
\left[
n_{\gamma}^{(+)}(\omega) n_{\gamma'}^{(+)}(\omega') n_{\gamma}''(\omega'') 
- n_{\gamma}(\omega) n_{\gamma'}(\omega') n_{\gamma''}^{(+)}(\omega'') 
\right]
 \delta (\hbar \omega +\hbar \omega' - \hbar \omega'')
 \Big\}
 \label{eq:pdist}
\end{eqnarray}
with $\gamma=$LA, TA1, and TA2. We introduced the coupling functions defined as
\begin{eqnarray}
%%%%%%%%%%%%%%%%
 C_{\rm e \mathchar`- e}(\varepsilon,\varepsilon', \xi,\xi') 
 &=& \frac{1}{\hbar {\cal N}(\varepsilon)}
 \sum_{\bm{k},\bm{k}',\bm{q}} \vert \tilde{V}(\bm{q}) \vert^2
 \delta (\varepsilon - \varepsilon_{\bm{k}})
  \delta (\varepsilon' - \varepsilon_{\bm{k}'})
   \delta (\xi - \varepsilon_{\bm{k}+\bm{q}})
    \delta (\xi' - \varepsilon_{\bm{k}'- \bm{q}}),
    \label{eq:coupling_e-e}
    \\
%%%%%%%%%%%%%%%%
 C_{\rm e \mathchar`- ph}(\varepsilon,\xi,\omega,\gamma) 
 &=& \frac{1}{\hbar {\cal N}(\varepsilon)}
 \sum_{\bm{k},\bm{Q}} \vert \tilde{g}(\bm{Q},\gamma) \vert^2
 \delta (\varepsilon - \varepsilon_{\bm{k}})
   \delta (\xi - \varepsilon_{\bm{k}+\bm{Q}})
    \delta (\omega - \omega_{\bm{Q}\gamma}),
    \label{eq:coupling_e-ph}
    \\
 %%%%%%%%%%%%%%%%
 C_{\rm ph \mathchar`- e}(\varepsilon,\xi, \omega,\gamma) 
 &=& \frac{1}{\hbar {\cal D}_\gamma(\omega)}
 \sum_{\bm{k},\bm{Q}} \vert \tilde{g}(\bm{Q},\gamma) \vert^2
 \delta (\varepsilon - \varepsilon_{\bm{k}})
   \delta (\xi - \varepsilon_{\bm{k}+\bm{Q}})
    \delta (\omega - \omega_{\bm{Q}\gamma}),
    \label{eq:coupling_ph-e}
    \\
%%%%%%%%%%%%%%%%%%%%%%%%%%%%%%%%
% Anharmonic and Inelastic coupling
%%%%%%%%%%%%%%%%%%%%%%%%%%%%%%%%     
C_{\rm ph \mathchar`- ph}(\omega, \omega', \omega'', \gamma, \gamma', \gamma'')
 &=& \frac{1}{\hbar {\cal D}_\gamma(\omega)}
 \sum_{\bm{Q},\bm{Q}'} 
 \vert \tilde{A}_{\bm{Q},\bm{Q}',-(\bm{Q} + \bm{Q}') }^{\gamma,\gamma',\gamma''} \vert^2
    \delta (\omega - \omega_{\bm{Q}\gamma})
        \delta (\omega' - \omega_{\bm{Q}'\gamma'})
            \delta (\omega'' - \omega_{\bm{Q} + \bm{Q}'\gamma''}).       
    \label{eq:coupling_ph-ph}
\end{eqnarray}
\end{widetext}
%%%%%%%%%%%%%%%%%%%%%%%%%%%%%%%%%%%%
For an isotropic system, these coupling functions are expressed as follows; The e-e coupling function in Eq.~(\ref{eq:coupling_e-e}),
\begin{eqnarray}
 & &C_{\rm e \mathchar`- e}(\varepsilon,\varepsilon'; \xi,\xi') 
 \nonumber\\
 &=& \frac{1}{4\pi^2 \varepsilon_{F}^{2}} \sqrt{\frac{\varepsilon_F}{\varepsilon}}
 \frac{\hbar}{m a_{0}^{2}} 
\int_{0}^{\infty} ds \left[ \frac{ds}{s^2 + (q_{\rm TF}/k_{\rm F})^2}\right]^2
\nonumber\\
&\times&
\theta_{\rm H} [1- h_1(\varepsilon,\xi,s)] \theta_{\rm H} [1+ h_1(\varepsilon,\xi,s)]
\nonumber\\
&\times&
\theta_{\rm H} [1- h_1(\varepsilon',\xi',s)] \theta_{\rm H} [1+ h_1(\varepsilon',\xi',s)]
\label{eq:Cee_iso}
\end{eqnarray}
with the Fermi wavenumber $k_{\rm F}$;
%%%%%%%%%%%%%%%%%
The e-ph coupling function in Eq.~(\ref{eq:coupling_e-ph}),
\begin{eqnarray}
 & &C_{\rm e \mathchar`- ph}(\varepsilon,\xi,\omega, \gamma) 
 \nonumber\\
 &=& \frac{3Z_{\rm val}}{128} \sqrt{\frac{\varepsilon_F}{\varepsilon}} 
 \left( \frac{D_\gamma}{\varepsilon_F}\right)^2
 \frac{(\hbar\omega)^2}{\left(\cfrac{1}{2}M_i v_{\gamma}^2\right)\left(\cfrac{1}{2}m v_{\gamma}^2\right)} 
\nonumber\\
&\times&
\theta_{\rm H} [1- h_2(\varepsilon,\xi,\omega)] \theta_{\rm H} [1+ h_2(\varepsilon,\xi,\omega)],
\label{eq:Cep_iso}
\end{eqnarray}
%%%%%%%%%%%%%%%%%
where $Z_{\rm val}$ is the number of the valence electron; and the ph-e coupling function in Eq.~(\ref{eq:coupling_ph-e}),
\begin{eqnarray}
 C_{\rm ph \mathchar`- e}(\varepsilon,\xi,\omega,\gamma) 
 &=& \frac{{\cal N}(\varepsilon)}{{\cal D}_\gamma (\omega)}
 C_{\rm e \mathchar`- ph}(\varepsilon,\xi,\omega,\gamma),
% \nonumber\\
% &=& (\frac{2\varepsilon_{\rm ac}}{\gamma r_s(\hbar\omega)^2} \frac{v_\gamma}{a_0} 
% \sqrt{\frac{\varepsilon}{\varepsilon_F}} 
% C_{\rm e \mathchar`- ph}(\varepsilon,\xi,\omega,\gamma) ), 
% \nonumber\\
\end{eqnarray}
%%%%%%%%%%%%%%%%%
where the functions $h_1$ and $h_2$ in Eqs.~(\ref{eq:Cee_iso}) and (\ref{eq:Cep_iso}) are given by
\begin{eqnarray}
h_1(\varepsilon,\xi,s)
&=& \frac{1}{2s} \sqrt{\frac{\varepsilon_F}{\varepsilon}} 
\left( \frac{\xi}{\varepsilon_F} - \frac{\varepsilon}{\varepsilon_F} - s^2 \right),
\\
h_2(\varepsilon,\xi,\omega)
&=& \sqrt{\frac{K_{\gamma} \varepsilon_{F}^{2}}{(\hbar\omega)^2\varepsilon}} 
\left[ \frac{\xi}{\varepsilon_F} - \frac{\varepsilon}{\varepsilon_F} 
- \frac{(\hbar\omega)^2}{4 K_{\gamma} \varepsilon_{\rm F}} \right]
\end{eqnarray}
with $K_{\gamma} = mv_{\gamma}^2 /2$. Note that when $\varepsilon =\xi = \varepsilon_F$, $C_{\rm e \mathchar`- ph}(\varepsilon,\xi,\omega,\gamma)$ is related to the Eliashberg function for the polarization $\gamma$ \cite{kabanov2008,baranov}
\begin{eqnarray}
 C_{\rm e \mathchar`- ph}(\varepsilon_F,\varepsilon_F,\omega,\gamma) 
 &=& \alpha^2F(\omega,\gamma)
\end{eqnarray}
with $\alpha^2F(\omega) = \sum_\gamma \alpha^2F(\omega,\gamma)$. Using this approximation, the e-ph coupling constants are defined by
\begin{eqnarray}
 \lambda_\gamma\langle \omega^n \rangle = 2 \int_{0}^{\Omega_{\gamma, {\rm D}}} d\omega 
 \alpha^2 F(\omega,\gamma) \omega^{n-1}
 \label{eq:lambdaomega2}
\end{eqnarray}
with $\lambda\langle \omega^n \rangle = \sum_\gamma \lambda_\gamma\langle \omega^n \rangle$. When the deviation of the single particle energy from the Femi energy is not negligible, we use the expression of
\begin{eqnarray}
 C_{\rm e \mathchar`- ph}(\varepsilon,\xi,\omega,\gamma) 
 &\simeq & \sqrt{\frac{\varepsilon_{\rm F}}{\varepsilon}}\alpha^2F(\omega,\gamma),
\end{eqnarray}
where the factor $\sqrt{\varepsilon_{\rm F} / \varepsilon}$ is multiplied [see the expression of Eq.~(\ref{eq:Cep_iso})], which will be used in Sec.~\ref{sec:NE+3T}.

%Within the free-electron approximation with the use of $D_0 = 2\varepsilon_F /3$,
%\begin{eqnarray}
%\gamma^2F(\omega)= \lambda \left( \frac{\hbar\omega}{2\hbar v_{\rm LA} k_F}\right)^2
% \theta(2\hbar v_{\rm LA} k_F - \hbar\omega)
%\end{eqnarray}
%with the dimensionless coupling constant
%\begin{eqnarray}
%\lambda = \frac{2Z_{\rm val}}{3} \frac{m}{M_i}\left( \frac{\varepsilon_F}{\hbar v_{\rm LA} k_F}\right)^2
%\end{eqnarray}
%where $Z_{\rm val}$ is the number of the valence electron.

The ph-ph coupling function satisfies, by definition, the following properties
\begin{eqnarray}
 & & C_{\rm ph \mathchar`- ph}(\omega,\omega',\omega'',\gamma,\gamma',\gamma'') 
 \nonumber\\
 &=& C_{\rm ph \mathchar`- ph}(\omega,\omega'',\omega',\gamma,\gamma'',\gamma')  
 \nonumber\\
  &=& \frac{{\cal D}_{\gamma'}(\omega')}{{\cal D}_{\gamma}(\omega)}C_{\rm ph \mathchar`- ph}(\omega',\omega,\omega'',\gamma',\gamma,\gamma'')  
\end{eqnarray}
The derivation of a simple expression for the ph-ph coupling function is difficult, so that we evaluate $C_{\rm ph \mathchar`- ph}$ in Eq.~(\ref{eq:coupling_ph-ph}) numerically. 

The details including the derivation of Eqs.~(\ref{eq:Cee_iso}) and (\ref{eq:Cep_iso}), and the numerical implementation of Eq.~(\ref{eq:coupling_ph-ph}) will be provided in the Appendix \ref{sec:app}.

%%%%%%%%%%%%%%%%%%%%%%%%%%%%%%%%%%%%
\subsubsection{NE+3T model}
\label{sec:NE+3T}
%%%%%%%%%%%%%%%%%%%%%%%%%%%%%%%%%%%%
Second, we present a model that consists of the nonequilibrium electrons and the quasiequilibrium phonons characterized by three phonon temperatures (NE+3T model). We assume that the effect of the e-e and ph-ph scatterings on the thermalization is negligible in the initial relaxation. Then, Eq.~(\ref{eq:edist}) is simplified into
\begin{eqnarray}
\frac{\partial f (\varepsilon)}{\partial t}
&=& 
2\pi \sum_{\gamma} \sum_{s= \pm} \int d\omega  
\sqrt{\frac{\varepsilon_{\rm F}}{\varepsilon}}
\alpha^2F(\omega,\gamma) 
U(\varepsilon,\omega,\gamma,s),
\nonumber\\
\end{eqnarray}
where
\begin{eqnarray}
U(\varepsilon,\omega,\gamma,s)
&=& 
[f(\varepsilon +s \hbar \omega) - f (\varepsilon)] n_{\rm BE}(\omega,T_{\rm ph}^{(\gamma)}) 
\nonumber\\
&+& s f (\varepsilon) [1- f(\varepsilon +s \hbar \omega )]
\end{eqnarray}
with the Bose-Einstein (BE) function $n_{\rm BE}(\omega,T_{\rm ph}^{(\gamma)})$. $T_{\rm ph}^{(\gamma)}$ is the phonon temperature for the polarization $\gamma$. Since the net phonon energy with $\gamma$ is defined by $E_{\rm ph, \gamma}^{\rm (net)} = \int d\omega \hbar\omega {\cal D}_\gamma (\omega) n_{\rm BE} (\omega,T_{\rm ph}^{(\gamma)})$, the $t$-evolution of $T_{\rm ph}^{(\gamma)}$ can be expressed as, by using Eq.~(\ref{eq:pdist}),
\begin{eqnarray}
C_{\rm ph}^{(\gamma)}
\frac{\partial T_{\rm ph}^{(\gamma)}}{\partial t}
&=& 4\pi N(\varepsilon_F) 
\int d\varepsilon  \int d\omega  \hbar\omega
\nonumber\\
&\times&
\sqrt{\frac{\varepsilon_{\rm F}}{\varepsilon}}\alpha^2 F(\omega,\gamma)
U(\varepsilon,\omega,\gamma,+),
\label{eq:NE+3T_ph}
\end{eqnarray}
where $C_{\rm ph}^{(\gamma)} = \partial E_{\rm ph, \gamma}^{\rm (net)}/\partial T_{\rm ph}^{(\gamma)}$ is the specific heat of the phonon with $\gamma$.

%%%%%%%%%%%%%%%%%%%%%%%%%%%%%%%%%%%%
\subsubsection{NP+1T model}
\label{sec:NP+1T}
%%%%%%%%%%%%%%%%%%%%%%%%%%%%%%%%%%%%
Third, we present a model that consists of the nonequilibrium phonons and the quasiequilibrium electrons (NP+1T model). The net electron energy is $E_{\rm e}^{\rm (net)} = 2\int d\varepsilon \varepsilon {\cal N} (\varepsilon) f_{\rm FD} (\varepsilon,T_{\rm e})$, where $T_{\rm e}$ is the electron temperature and $ f_{\rm FD}$ is the Fermi-Dirac (FD) function. The $t$-evolution of $E_{\rm e}^{\rm (net)}$ is expressed by 
\begin{eqnarray}
\frac{\partial E_{\rm e}^{\rm (net)}}{\partial t}
&=& - 4\pi N(\varepsilon_F)
\sum_\gamma \int d\omega \alpha^2 F(\omega,\gamma) (\hbar\omega)^2 
\nu_{\gamma }(\omega,T_{\rm e}),
\label{eq:Ee}
\nonumber\\
\end{eqnarray}
where 
\begin{eqnarray}
\nu_{\gamma}(\omega,T_{\rm e}) &=& n_{\rm BE} (\omega, T_{\rm e}) - n_\gamma (\omega).
\label{eq:nu_j}
\end{eqnarray}
In the derivation of Eq.~(\ref{eq:Ee}), we assumed that the e-ph coupling function is approximated to the Eliashberg function $\alpha^2 F (\omega,\gamma)$. Again, the contribution from the e-e and ph-ph scatterings are omitted in Eq.~(\ref{eq:Ee}). Thus, Eq.~(\ref{eq:Ee}) is a natural extension of Eq.~(10) in Ref.~\cite{allen}, since the formula is appropriate for the presence of nonequiibrium phonons. Using the Sommerfeld expansion \cite{ashcroft_mermin}, $E_{\rm e}^{\rm (net)}$ is expressed by the $t$-dependent temperature $T_{\rm e}(t)$, that is, $E_{\rm e}^{\rm (net)}(t) = 3N_{\rm e} \varepsilon_F /5 + \pi^2 N(\varepsilon_F)\left[k_{\rm B}T_{\rm e}(t)\right]^2/3$ with the total number of the electrons $N_{\rm e}$. Then, one obtains
\begin{eqnarray}
\frac{\partial T_{\rm e}}{\partial t}
&=& - \frac{6}{\pi k_{B}^{2} T_{\rm e}}
\sum_\gamma \int d\omega \alpha^2 F(\omega,\gamma) (\hbar\omega)^2 
\nu_{\gamma}(\omega,T_{\rm e}).
\nonumber\\
\label{eq:Te_rate2}
\end{eqnarray}
Using Eq.~(\ref{eq:pdist}), the $t$-evolution of $n_{\gamma}(\omega)$ is simply written as 
\begin{eqnarray}
\frac{\partial n_{\gamma} (\omega)}{\partial t}
&=& \frac{4\pi N(\varepsilon_F) \hbar\omega }{D_\gamma (\omega)}
\alpha^2 F(\omega,\gamma) 
\nu_{\gamma}(\omega,T_{\rm e}).
%\nonumber\\
%&\equiv& {\cal A}_j \nu_{j}(\omega,T_{\rm e})
\label{eq:nj_2}
\end{eqnarray}
%It should be emphasized that the quasiequilibrium of phonons is not assumed in this model.

%%%%%%%%%%%%%%%%%%%%%%%%%%%%%%%%%%%%
\subsubsection{4TM}
\label{sec:4TM}
%%%%%%%%%%%%%%%%%%%%%%%%%%%%%%%%%%%%
To further simplify the model, we assume that the electron and each phonon subset are also quasiequilibrium. We replace $f(\varepsilon)$ in Eq.~(\ref{eq:NE+3T_ph}) and $n_\gamma (\omega)$ in Eq.~(\ref{eq:Te_rate2}) with with $f_{\rm FD} (\varepsilon,T_{\rm e})$ and $n_{\rm BE} (\omega,T_{\rm ph}^{(\gamma)})$, respectively. Using the high temperature approximation $\hbar\omega/(k_{\rm B}T) \ll 1$, that is, $n_{\rm BE} \simeq k_{\rm B}T/(\hbar\omega)$, one obtains the 4TM
\begin{eqnarray}
 C_{\rm e} \frac{\partial T_{\rm e}}{\partial t} 
 &=& - \sum_\gamma g_{\gamma} (T_{\rm e} -T_{\rm ph}^{(\gamma)}),
 \nonumber\\
 C_{\rm ph}^{(\gamma)} \frac{\partial T_{\rm ph}^{(\gamma)}}{\partial t} 
 &=& g_{\gamma} (T_{\rm e} -T_{\rm ph}^{(\gamma)}),
%+ \sum_j G_{ij} (T_{ph}^{(j)} -T_{ph}^{(i)})
%\nonumber\\
\end{eqnarray}
where $C_{\rm e}$ is the specific heat of the electron. The coefficients $g_\gamma$ is written as 
\begin{eqnarray}
\frac{g_\gamma}{C_{\rm e}} 
=
\frac{ 3\hbar \lambda_\gamma\langle \omega^2 \rangle} {\pi k_{\rm B}T_{\rm e}}.
\end{eqnarray}

%%%%%%%%%%%%%%%%%%%%%%%%%%%%%%%%%%%%
\subsubsection{2TM}
\label{sec:2TM}
%%%%%%%%%%%%%%%%%%%%%%%%%%%%%%%%%%%%
In the original derivation of the 2TM by Allen \cite{allen}, all the phonon temperatures are assumed to be the same. Then, one obtains
\begin{eqnarray}
 C_{\rm e} \frac{\partial T_{\rm e}}{\partial t} &=& - G (T_{\rm e} -T_{\rm ph}),
 \nonumber\\
 C_{\rm ph} \frac{\partial T_{\rm ph}}{\partial t} &=& G (T_{\rm e} -T_{\rm ph}),
\end{eqnarray}
where $G = \sum_{\gamma} g_\gamma$. The important fact is that the coefficient $G$ in the 2TM contains the e-ph coupling constant $\lambda \langle \omega^2 \rangle = \sum_\gamma\lambda_\gamma\langle \omega^2 \rangle$. The measurement of $T_{\rm e}$ as a function of $t$ will make it possible to determine the magnitude of $\lambda \langle \omega^2 \rangle$, if many assumptions above, which might be difficult to be satisfied, are correct.

%%%%%%%%%%%%%%%%%
\begin{table}[bbb]
\begin{center}
\caption{Material parameters}
{
\begin{tabular}{ll}\hline
%--------------------------------------------------------------------------------
$r_{s}$ & Wigner-Seitz radius  \\ 
$\lambda_{\rm L}$, $\mu_{\rm L}$ & The second-order elastic constants \\ 
$E_1, E_2, E_3$ & The third-order elastic constants \\ 
%--------------------------------------------------------------------------------
\hline
\end{tabular}
}
\label{tab:parameters}
\end{center}
\end{table}
%%%%%%%%%%%%%%%%%%%%%%%%%%%%%%%%%%%%%%%%%%%%%%%%%%%

%%%%%%%%%%%%%%%%%
\begin{table}[bbb]
\begin{center}
\caption{The second- and third-order elastic constants of Al \cite{hwang} in units of GPa.}
{
\begin{tabular}{llllllll}\hline
%--------------------------------------------------------------------------------
$C_{11}$ & $110.4$ & \hspace{2mm} &
$C_{12}$ & $54.5$ & \hspace{2mm} &
$C_{44}$ & $31.3$ \\
$C_{111}$ & $-1253 $ & \hspace{2mm} &
$C_{112}$ & $ -426 $ & \hspace{2mm} &
$C_{123}$ & $ 153 $  \\ 
$C_{144}$ & $ -12 $ & \hspace{2mm} &
$C_{166}$ & $-493 $ & \hspace{2mm} &
$C_{456}$ & $-21 $  \\ 
%--------------------------------------------------------------------------------
\hline
\end{tabular}
}
\label{tab:Al_parameters}
\end{center}
\end{table}
%%%%%%%%%%%%%%%%%%%%%%%%%%%%%%%%%%%%%%%%%%%%%%%%%%%

%%%%%%%%%%%%%%%%%%%%%%%%%%%%%%%%%%%%
\subsection{Computational details}
In the present model, there are six parameters that describe the material properties. These are listed in Table \ref{tab:parameters}. In the present study, we study the thermalization of aluminum (Al), the most studied simple metal. Thus, we set $r_{s}=2.07$ \cite{ashcroft_mermin}. To determine the elastic constants of the isotropic system from those of a real solid, we minimize $X_2$ and $X_3$ defined as \cite{tamura}
\begin{eqnarray}
 X_2 &=& \sum_{ijkl} (G_{ijkl} - C_{ijkl})^2,
 \nonumber\\
  X_3 &=& \sum_{ijklmn} (G_{ijklmn} - C_{ijklmn})^2,
\end{eqnarray}
where $G_{ijkl}$ and $G_{ijklmn}$ are the elastic constants of a realistic system. For cubic crystals, one obtains
\begin{eqnarray}
 \lambda_{\rm L} &=& \frac{1}{5} (C_{11}+4C_{12}-2C_{44}),
 \nonumber\\
 \mu_{\rm L} &=& \frac{1}{5} (C_{11}-C_{12}+3C_{44}),
 \nonumber\\
 E_{1} &=& \frac{1}{35} \, C_{111} + \frac{18}{35} \, C_{112} + \frac{16}{35} \, C_{123} 
 \nonumber\\
 &-& \frac{6}{7} \, C_{144} - \frac{12}{35} \, C_{244} + \frac{16}{35} \, C_{456}, 
 \nonumber\\
 E_{2} &=& \frac{1}{35} \, C_{111} + \frac{4}{35} \, C_{112} - \frac{1}{7} \, C_{123} 
 \nonumber\\
 &+& \frac{19}{35} \, C_{144} + \frac{2}{35} \, C_{244} - \frac{12}{35} \, C_{456}, 
 \nonumber\\
 E_{3} &=& \frac{1}{35} \, C_{111} - \frac{3}{35} \, C_{112} + \frac{2}{35} \, C_{123} 
 \nonumber\\
 &-& \frac{9}{35} \, C_{144} + \frac{9}{35} \, C_{244} + \frac{9}{35} \, C_{456},
 \label{eq:elastic_constants}
\end{eqnarray}
where the Voigt notation $(11\rightarrow 1, 22\rightarrow2, 33\rightarrow3, 12\rightarrow4, 23\rightarrow5, 31\rightarrow6)$ is used. The values of $C_{ij}$ and $C_{ijk}$ in the right-hand side of Eq.~(\ref{eq:elastic_constants}) have been computed from density-functional theory calculations \cite{hwang}. Using the values listed in Table \ref{tab:Al_parameters}, we obtain $\lambda_{\rm L}$, $\mu_{\rm L}$, $E_1$, $E_2$, and $E_3$. 

To determine the magnitude of the deformation potential for the TA phonons, $D_{\rm TA}$, introduced phenomenologically below Eq.~(\ref{eq:eph_matrix}), we set $\beta = 1.5$. The use of Eq.~(\ref{eq:lambdaomega2}) yields $\lambda = 0.383$ ($\lambda_{\rm LA} = 0.191$ and $\lambda_{\rm TA} = 0.096$) and $\lambda \langle \omega^2 \rangle = 510.0$ meV$^2$ ($\lambda_{\rm LA} \langle \omega^2 \rangle = 404.9$ meV$^2$ and $\lambda_{\rm TA} \langle \omega^2 \rangle = 52.5$ meV$^2$), which are similar values reported in Al \cite{lin2008,grimvall}.

The BTE given by Eqs.~(\ref{eq:edist}) and (\ref{eq:pdist}) is solved by the fourth-order Runge-Kutta method, setting the time step of 0.2419 fs. In the present model, the Fermi energy is $\varepsilon_{\rm F}=11.695$ eV and the Debye energy for the LA phonon is $E_0 = \hbar\Omega_{\rm LA,D}=65$ meV. Within an electron energy window of $\varepsilon \in \left[\varepsilon_{\rm F} - 10E_0,\varepsilon_{\rm F}+10E_0\right]$, 1600 discrete electron energies are considered. On one hand, 80 and 41 discrete phonon energies for the LA and TA modes are considered. 

%%%%%%%%%%%%%%%%%%%%%%%%%%%%%%%%%%%%
\section{Results and Discussion}
\label{sec:results}
Using $f(\varepsilon)$ and $n_\gamma(\omega)$ at $t$, the excess electron and phonon energies, measured from the total energies in thermal equilibrium, are computed by 
\begin{eqnarray}
E_{\rm e}
&=& 
2\int d\varepsilon \varepsilon N(\varepsilon) [f(\varepsilon) - f_{\rm FD}(\varepsilon,T_0)],
\label{eq:totE_e}
\\
E_{{\rm ph},\gamma}
&=& 
\int d\omega \hbar\omega D_\gamma(\omega) [n_\gamma(\omega) - n_{\rm BE}(\omega,T_0)].
\label{eq:totE_ph}
\end{eqnarray}
The factor of $2$ in Eq.~(\ref{eq:totE_e}) comes from the spin degeneracy. In the present study, we set $T_0=25$ meV.

It should be noted that the effect of the e-e scattering on the electron thermalization was found to be negligibly small, that is, the $t$-$E_{\rm e}$ curves with and without the e-e scattering effect almost overlap. This may be due to a relatively small $r_s$ of Al, yielding a strong screening effect [see Eqs.~(\ref{eq:thomas}) and (\ref{eq:scp})]. Thus, we will not consider the impact of the e-e scatterings explicitly below. 

%%%%%%%%%%%%%%%%%%%%%%%%%%%%%%%%%%%%%%%
\subsection{Thermalization}
\label{sec:thermalization}
As a model study, we start from the initial electron distribution function with Gaussian-type peaks above and below the Fermi level, which is given by
\begin{eqnarray}
f(\varepsilon) = f_{\rm FD}(\varepsilon,T_0) 
+ \sum_{s=\pm} sA_s \exp\left[
-\left(\frac{\varepsilon-s\varepsilon_0}{2W}\right)^2
\right],
\label{eq:excitation}
\end{eqnarray}
where $\varepsilon_0$ and $W$ are the peak position and the width, respectively. Given a peak height $A_{-}$, $A_+$ is uniquely determined by the electron number conservation law. The phonon distribution is assumed to be $n_{\rm BE}(\omega,T_0)$. Similar types of the model functions have been used to study the thermalization in metals \cite{groeneveld}. In this subsection, we set $A_{-}=0.1$, $\varepsilon_0=300$ meV, and $W=50$ meV, where the effective electron temperature amounts to approximately 800 K. In Sec.~\ref{sec:noneq}, the effect of the initial distribution on the thermalization will be investigated.

%%%%%%%%%%%%%%%%%
\begin{figure}[t]
\center
\includegraphics[scale=0.45,clip]{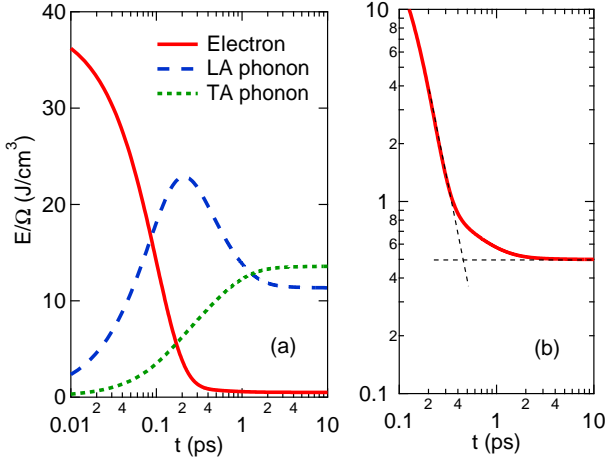}
\caption{\label{fig:1} (a) The $t$-dependence of $E_{\rm e}$, $E_{\rm ph,LA}$, and $E_{\rm ph,TA}(=E_{\rm ph,TA1}=E_{\rm ph,TA2})$ computed by Eqs.~(\ref{eq:totE_e}) and (\ref{eq:totE_ph}). (b) The magnified view of $t$-$E_{\rm e}$ curve below $E_{\rm e}=10$ J/cm$^3$ in a log-log plot. }
\end{figure}
%%%%%%%%%%%%%%%%%
%%%%%%%%%%%%%%%%%
\begin{figure}[t]
\center
\includegraphics[scale=0.45,clip]{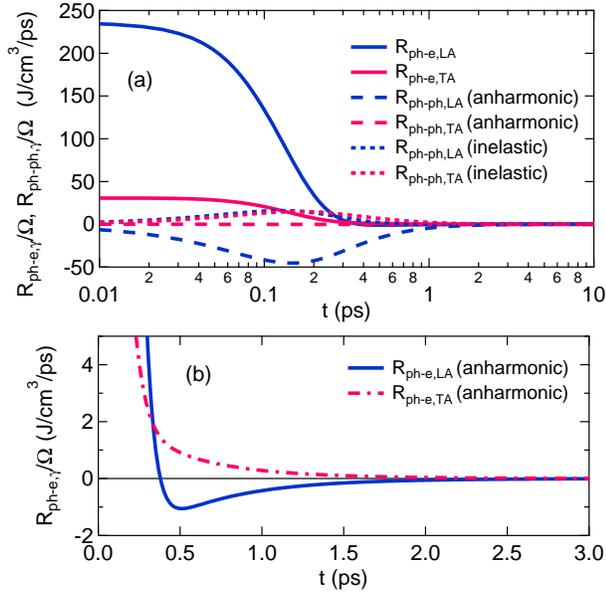}
\caption{\label{fig:2}(a) The $t$-dependence of $R_{{\rm ph \mathchar`- e}, \gamma}$ and $R_{{\rm ph \mathchar`- ph}, \gamma}$ with $\gamma=$LA and TA1 computed from Eqs.~(\ref{eq:R_ph_e}) and (\ref{eq:R_ph_ph}). For the inelastic scatterings, the curves of $R_{\rm ph \mathchar`- ph, LA}$ and $R_{\rm ph \mathchar`- ph, TA}$ almost overlap. (b) The magnified view of $R_{{\rm ph \mathchar`- e}, \gamma}$ up to $t=3$ ps.}
\end{figure}
%%%%%%%%%%%%%%%%%

%During the relaxation dynamics, the electrons are quasiequilibrium, while the phonons are nonequilibrium; the electron distribution can be well described by the Fermi distribution function with the electron temperature $T_{e}$. Figure \ref{fig:1} shows the $t$-dependence of $T_{e}$. $T_e$ decreases with increasing $t$, indicating the energy transfer to the phonons. The thermalization process is decomposed into three stages: (i) $t\in [0,0.2]$ ps, (ii) $t\in [0.2,5]$ ps, and (iii) $t \in [5, \infty)$ ps. In the second stage of the relaxation (ii), $T_{e}$ decays $\propto t^{-\gamma}$ with $\gamma = 0.14$. The origin of the power-law decay will be discussed below. 
%To understand the three thermalization processes above, 

Figure \ref{fig:1}(a) shows the $t$-evolution of $E_{\rm e}$, $E_{\rm ph, LA}$, and $E_{\rm ph, TA1}$ defined as Eqs.~(\ref{eq:totE_e}) and (\ref{eq:totE_ph}). Since the contribution from $E_{\rm ph, TA2}$ is the same as $E_{\rm ph, TA1}$, the former is not shown in Fig.~\ref{fig:1}(a). Hereafter, we denote $E_{\rm ph, TA1}$ and $E_{\rm ph, TA2}$ as $E_{\rm ph, TA}$. In the initial stage of the relaxation, $t\le 0.2$ ps, $E_{\rm e}$ decreases drastically, while $E_{\rm ph, LA}$ and $E_{\rm ph, TA}$ gradually increases. Most of the electron energy is transferred into the LA phonons, resulting in the overshoot of $E_{\rm ph, LA}$ at $t=\tau_0 = 0.21$ ps, followed by a slow decay of $E_{\rm ph, LA}$ and a slow increase in the $E_{\rm ph, TA}$. After $t=3$ ps, the values of $E_{\rm el}$, $E_{\rm ph, LA}$, and $E_{\rm ph, TA}$ are almost constant, indicating the thermal equilibrium of the system. 

To understand the overshoot of the LA phonon energy, we decompose the energy transfer rate $\partial E_{{\rm ph},\gamma}/\partial t$ into a sum of the contribution from the ph-e and ph-ph scatterings, which is defined as, respectively,
\begin{eqnarray}
 R_{{\rm ph \mathchar`- e}, \gamma} 
 &=& \left(\frac{\partial E_{{\rm ph},\gamma}}{\partial t}\right)_{\rm ph \mathchar`- e}, 
 \label{eq:R_ph_e}
 \\
  R_{{\rm ph \mathchar`- ph}, \gamma} 
  &=& \left(\frac{\partial E_{{\rm ph},\gamma}}{\partial t}\right)_{\rm ph \mathchar`- ph}.
\label{eq:R_ph_ph}
\end{eqnarray}
The latter, Eq.~(\ref{eq:R_ph_ph}), is further decomposed into two contributions: the anharmonic decay $R_{{\rm ph \mathchar`- ph}, \gamma}{\rm (anharmonic)}$ and the inelastic scatterings $R_{{\rm ph \mathchar`- ph}, \gamma}{\rm (inelastic)}$. Similarly, we can define the energy transfer rate of the electronic system due to the e-ph scatterings, but the quantity is exactly equal to $- R_{{\rm ph \mathchar`- e}, \gamma}$. 

Figure \ref{fig:2}(a) shows $R_{{\rm ph \mathchar`- e},\gamma}$ and $R_{{\rm ph \mathchar`- ph}, \gamma}$ as a function of $t$. Within the initial relaxation ($t\ll \tau_0$ ps), $R_{\rm ph \mathchar`- e, LA}$ is positive and much larger than the other rates, indicating that the LA phonons obtain a large amount of the electron energy via the ph-e scatterings. Simultaneously, the magnitude of $\vert R_{\rm ph \mathchar`- ph, LA}{\rm (anharmonic)}\vert$ increases with increasing $t$. Negative value of $R_{\rm ph \mathchar`- ph, LA}{\rm (anharmonic)}$ indicates that the LA phonon decays into low-frequency LA and TA phonons. Correspondingly, both of $R_{\rm ph \mathchar`- ph, LA}{\rm (inelastic)}$ $R_{\rm ph \mathchar`- ph, TA}{\rm (inelastic)}$ increase with time. The sum of the contribution from the anharmonic and the inelastic processes, $\vert R_{\rm ph \mathchar`- ph, LA}\vert$, becomes larger than the value of $R_{\rm ph \mathchar`- e, LA}$ after $t = \tau_0$ ps. In this way, the overshoot of $E_{\rm ph, LA}$, shown in Fig.~\ref{fig:1}(a), is due to a crossover from the energy gain via the ph-e scattering to the energy loss via the anharmonic decay into TA phonons. 

Figure \ref{fig:1}(b) is a magnified view of $E_{\rm e}$ shown in Fig.~\ref{fig:1}(a). Before $t\simeq 0.4$ ps, $E_{\rm e}$ decreases linearly in a log-log plot, while before reaching the thermal equilibrium ($t = 3$ ps) the relaxation behavior slows. Figure \ref{fig:2}(b) also shows a magnified view of $R_{\rm ph \mathchar`- e, LA}$ and $R_{\rm ph \mathchar`- e, TA}$ in Fig.~\ref{fig:1}(b). Negative value of $R_{\rm ph \mathchar`- e, LA}$ is observed during the time interval $t\in [0.38,4.7]$ ps. This means that the energy stored in the LA phonons is transferred, in turn, into the electronic system. The time interval showing $R_{\rm ph \mathchar`- e, LA} <0$ is almost the same as the interval when the electron relaxation slows [Fig.~\ref{fig:1}(b)]. Thus the appearance of the backward energy flow is an indicator for reaching a thermal equilibrium. Similar scenario involving the backward energy flow has been reported in phononic systems \cite{ono2017}. Note that the negative $R_{\rm ph \mathchar`- e, LA}$ is also observed after $t=8.7$ ps, while the value is quite small: $R_{\rm ph \mathchar`- e, LA}\sim -10^{-4}$ J/(cm$^3\cdot$ps). This is similar to the dynamics of the damped oscillators in the classical mechanics when $R_{\rm ph \mathchar`- e, LA}$ in Eq.~(\ref{eq:R_ph_e}) is regarded as the amplitude of the oscillation.

%%%%%%%%%%%%%%%%%
\begin{figure}[t]
\center
\includegraphics[scale=0.4,clip]{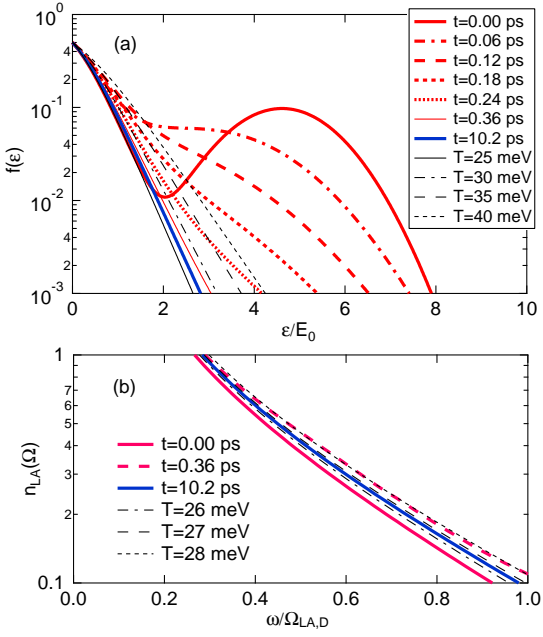}
\caption{\label{fig:3} (a) $f(\varepsilon)$ and (b) $n_{\rm LA} (\omega)$ for various $t$s. The FD and BE functions with several $T$s are also shown. $f(\varepsilon)$ is nearly an odd function with respect to $\varepsilon=0$.}
\end{figure}
%%%%%%%%%%%%%%%%%
%%%%%%%%%%%%%%%%%
\begin{figure}[t]
\center
\includegraphics[scale=0.4,clip]{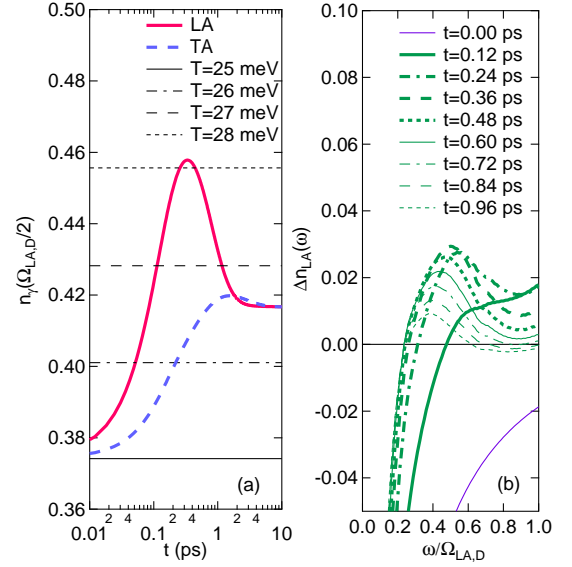}
\caption{\label{fig:4} (a) $n_{\rm LA} (\Omega_{\rm LA,D}/2)$ and $n_{\rm TA} (\Omega_{\rm LA,D}/2)$ as a function of $t$. The values of $n_{\rm BE}(\Omega_{\rm LA,D}/2,T)$ for $T=25,26,27,28$ meV are shown. (b) The deviation of $n_{\rm LA} (\omega)$ from the BE function with $T=27$ meV for several $t$s. }
\end{figure}
%%%%%%%%%%%%%%%%%

The decay process discussed above could be understood in terms of the nonequilibrium distribution functions. Figures~\ref{fig:3}(a) and (b) show $f(\varepsilon)$ and $n_{\rm LA}(\omega)$ for various $t$s, computed by Eqs.~(\ref{eq:edist}) and (\ref{eq:pdist}), respectively. For comparison, the FD and BE distribution functions with several $T$s are also shown. The Gaussian peak observed at $\varepsilon/E_0 =5$ in $f(\varepsilon)$ is immediately smeared out within 0.2 ps, while the deviation from the FD function is still not negligible, which is similar to the numerical results in Ref.~\cite{baranov}. After $t=0.36$ ps, the quasiequilibrium treatment for $f(\varepsilon)$ may be valid, shown in Fig.~\ref{fig:3}(a). In contrast, the LA phonon population increases due to the ph-e scatterings up to $t=0.36$ ps, after which they decreases, as shown in Fig.~\ref{fig:3}(b). To understand the phonon dynamics quantitatively, we show the $t$-dependence of $n_\gamma (\omega)$ at $\omega = \Omega_{\rm LA,D}/2$ in Fig.~\ref{fig:4}(a). Quite similarly to $t$-$E_{\rm ph, LA}$ curve in Fig.~\ref{fig:1}(a), an overshoot of the LA phonon population is observed around $t=0.34$ ps, while the change in $n_{\rm TA} (\Omega_{\rm LA,D}/2)$ is moderate. Figure~\ref{fig:4}(b) shows the deviation of $n_{\rm LA} (\omega)$ from the BE statistics at $T=27$ meV for several $t$s. The phonon population at $\omega = \Omega_{\rm LA,D}$ increases with time and is maximum at $t=0.12$ ps, after which it decreases. Since the ph-ph scatterings such as LA $\leftrightarrows$ LA + TA frequently occur after the creation of the hot high-frequency LA phonon, the population of the LA phonon with lower frequency ($\omega \simeq \Omega_{\rm LA,D}/2$) increases with time. 

The observation of $R_{\rm ph \mathchar`- e, LA} =0$ at $t = 0.38$ ps, in Fig.~\ref{fig:2}(b), is closely related to the breakdown of the quasiequilibrium approximation for the LA phonon. Given the quasiequilibrium distribution for $f(\varepsilon)$ after $t=0.38$ ps, the value of $R_{{\rm ph \mathchar`- e},\gamma}$ can be approximated as
\begin{eqnarray}
R_{{\rm ph \mathchar`- e},\gamma}
&\simeq& 4\pi N(\varepsilon_{\rm F}) 
\int d\omega \alpha^2 F(\omega,\gamma) 
\nu_\gamma (\omega,T_{\rm e}).
\end{eqnarray}
A trivial solution to $R_{{\rm ph \mathchar`- e},\gamma}=0$ is $n_{\gamma}(\omega) = n_{\rm BE}(\omega,T_{\rm e})$, that is, a thermal equilibrium at $t\rightarrow \infty$. Another solution is $n^{0}_{\gamma}(\omega)$ satisfying a relation
\begin{eqnarray}
\int d\omega \alpha^2 F(\omega,\gamma) n^{0}_{\gamma} (\omega) = 
\int d\omega \alpha^2 F(\omega,\gamma) n_{\rm BE}(\omega,T_{\rm e}).
\label{eq:balance}
\end{eqnarray}
Since $n_{\gamma}(\omega)$ at $t=0.38$ ps deviates from the BE function positively for higher $\omega$ and negatively for lower $\omega$ shown in Fig.~\ref{fig:4}(b), Eq.~(\ref{eq:balance}) holds. This clearly shows the breakdown of the quasiequilibrium treatment for LA phonons.

%%%%%%%%%%%%%%%%%
\begin{figure}[t]
\center
\includegraphics[scale=0.45,clip]{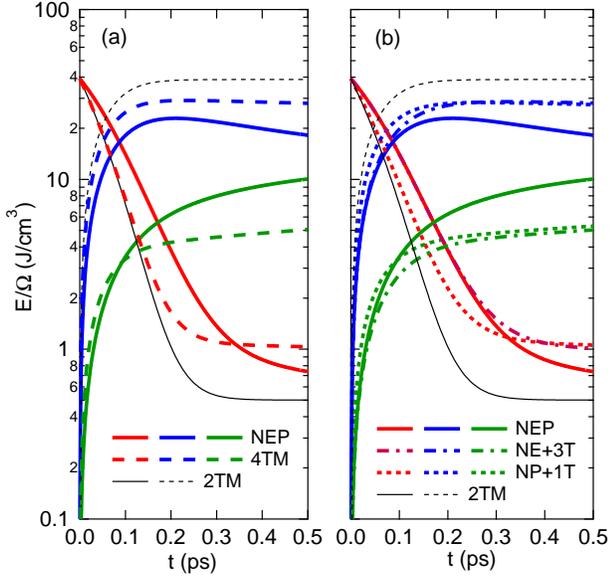}
\caption{\label{fig:5} The $t$-dependence of the excess energies of the electrons and phonons computed within (a) the NEP model, 4TM, and 2TM and (b) the NEP, NE+3T and NP+1T models, and 2TM. The deviations from the NEP model in $E_{\rm e}$ for $t > 0.3$ ps and $E_{{\rm ph,} \gamma}$ for $t > 0.1$ ps are due to the lack of the ph-ph scattering effect in the NE+3T and NP+1T models.}
\end{figure}
%%%%%%%%%%%%%%%%%

%%%%%%%%%%%%%%%%%
\begin{figure}[t]
\center
\includegraphics[scale=0.55,clip]{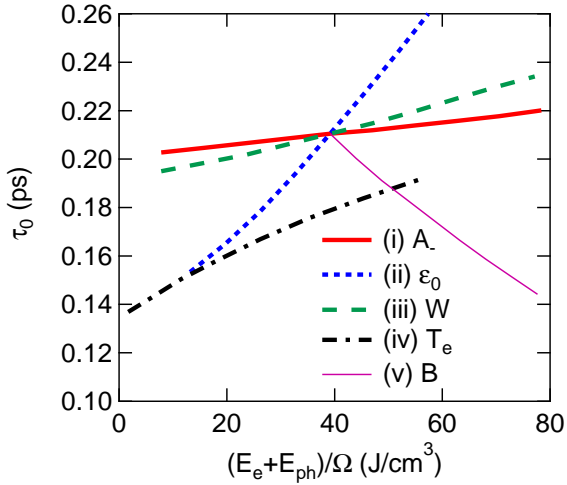}
\caption{\label{fig:6} $\tau_0$ as a function of $E_{\rm e}(t=0) + E_{\rm ph}(t=0)$ for various $f(\varepsilon,t=0)$s in Eq.~(\ref{eq:excitation}) and $n_{\rm LA}(\omega,t=0)$s in Eq.~(\ref{eq:excitation_ph}): From the case of (i) to (v). See text for the details.}
\end{figure}
%%%%%%%%%%%%%%%%%

%%%%%%%%%%%%%%%%%%%%%%%%%%%%%%%%%%%%%%%
\subsection{Impact of the nonequilibrium distribution}
\label{sec:noneq}
The total energy dynamics within the 4TM and 2TM, described in Secs.~\ref{sec:4TM} and \ref{sec:2TM}, respectively, are shown in Figs.~\ref{fig:5}(a). The $t$-$E_{\rm e}$, $E_{\rm ph, LA}$, and $E_{\rm ph, TA}$ curves in the NEP model are also shown. Clearly, the initial electron relaxation derived from the 2TM and 4TM is faster than that derived from the NEP model. Figure \ref{fig:5}(b) shows the relaxation dynamics in the NE+3T and NP+1T models described in Secs.~\ref{sec:NE+3T} and \ref{sec:NP+1T}, respectively. The former model reproduces the initial relaxation in the NEP model, while the latter model improves the relaxation behavior slightly, compared with the 2TM. What is important in these comparative studies is that the energy relaxation in the quasiequilibrium treatment is faster than that in the nonequilibrium treatment. This would lead to an underestimation of the e-ph coupling constant when the 2TM or 4TM are applied to time-resolved experiments. This result, the underestimation of $\lambda\langle \omega^2 \rangle$, is consistent with the results reported in Ref.~\cite{waldecker}.

We define $\tau_0$, at which $E_{\rm ph, LA}$ is the maximum, as the initial relaxation time, which may be determined from time-resolved diffraction experiments \cite{trigo,harb}. Figure \ref{fig:6} shows $\tau_0$ as a function of $E_{\rm e}(t=0)$ for various $f(\varepsilon,t=0)$. In Eq.~(\ref{eq:excitation}), we tune three parameters as follows; (i) $A_{-} \in [0.02,0.2]$, $\varepsilon_0 =300$ meV, and $W=50$ meV; (ii) $A_{-}=0.1$, $\varepsilon_0 \in [100,450]$ meV, and $W = 50$ meV; and (iii) $A_{-} =0.1$, $\varepsilon_0 =300$ meV, and $W \in [10,100]$ meV. We also studied the case of (iv) $f(\varepsilon,t=0)$ equal to the quasiequilibrium distribution with $T_{\rm e}(t=0) \in [30,100]$ meV. As expected, $\tau_0$ increases with increasing $E_{\rm e}(t=0)$ for all cases (i)-(iv). However, the curves of $\tau_0$-$E_{\rm e}(t=0)$ cannot be described by a single function, implying that $\tau_0$ is {\it a functional} of $f(\varepsilon,t=0)$. Unfortunately, we could not find a clear relationship between them, whose derivation or microscopic understanding will be a future work. As shown in Fig.~\ref{fig:6}, the quasiequilibrium approximation, the case (iv), gives the lowest value of $\tau_0$. Similar result has been reported in Ref.~\cite{rethfeld2002}, while the initial distribution used is different from that used in the present study.

More realistically, the electron distribution will change due to the electron-photon interaction within a pulse width \cite{rethfeld2002,mueller}. During the finite pulse width, the phonons are also created through the ph-e scatterings, disturbing the phonon distribution function. To consider this effect, we assume that the initial LA phonon distribution is given as
\begin{eqnarray}
 n_{\rm LA}(\omega) = n_{\rm BE}(\omega,T_0) 
 + B \exp \left[ -\left(\frac{\omega - \Omega_{\rm LA,D}}{2W_{\rm ph}/\hbar}\right)^2 \right],
 \label{eq:excitation_ph}
\end{eqnarray}
where $B$ and $W_{\rm ph}$ are the peak height and the width. $n_{\rm TA}(\omega)$ is assumed to be equal to $n_{\rm BE}(\omega,T_0)$ at $t=0$ ps because the magnitude of the electron-TA phonon coupling is smaller than that of the electron-LA phonon coupling. The initial electron distribution is still expressed as Eq.~(\ref{eq:excitation}). We set $A_{-}=0.1$, $\varepsilon_0 =300$ meV, $W=50$ meV, $B \in [0,0.16]$, and $W_{\rm ph} = 0.1 E_0$, as the case of (v). Nonzero $B$ yields a positive value of $E_{\rm ph}(t=0)$. $\tau_0$ as a function of $E_{\rm e}(t=0)+E_{\rm ph}(t=0)$ is shown in Fig.~\ref{fig:6}. Contrary to the cases of (i)-(iv), $\tau_0$ decreases as the sum of the excess energies increases. This means that the anharmonic decay event occurs frequently with increasing $E_{\rm ph, LA}(t=0)$, as a result of which $\tau_0$ becomes faster. What is suggested by this special condition is that the phonon population dynamics would play an important role in understanding the thermalization in metals. Note that the overshoot of $E_{\rm ph, LA}$ will vanish if $E_{\rm ph, LA}(t=0)/E_{\rm e}(t=0) \gg 1$.

%%%%%%%%%%%%%%%%%%%%%%%%%%%%%%%%%%%%
\section{Summary}
\label{sec:summary}
We have studied the electron and phonon thermalization in photoexcited metals by solving the BTE taking into account the e-e, e-ph, ph-e, and ph-ph scatterings. We have found that in the initial stage of the relaxation, most of the electron energy is transferred into the LA phonons through the e-ph and ph-e scatterings. Simultaneously, the LA phonon decays into TA phonons via the ph-ph scatterings. This yields an overshoot of the total LA phonon energy at a time $\tau_0$. The behavior of the thermalization is not affected by the presence of the e-e scattering. The picture for the thermalization demonstrated in the present study is quite different from the 2TM scenario \cite{allen}. 

By systematically investigating the relaxation dynamics of several models, we have shown that the energy relaxation within the quasiequilibrium approximation is found to be faster than that in a realistic situation. This implies that the use of the 2TM underestimates the e-ph coupling constant in metals, consistent with the results in Ref.~\cite{waldecker}.

We have also found that the relaxation time $\tau_0$ strongly depends on the initial distribution functions $f(\varepsilon,t=0)$ and $n_\gamma(\omega,t=0)$ as well as the initial energy. An open question is to find a functional form to estimate $\tau_0$. Such a functional must also contain the information about the e-ph and ph-ph interactions. 

%Several models, different from the 2TM and 4TM, describing the nonequilibrium electron-phonon dynamics of solids have been developed so far, for semiconductors \cite{sadasivam,brouwer2017} and superconductors \cite{RT,kabanov1999,unter2008,ono2012}. 

\begin{acknowledgments}
This study is supported by a Grant-in-Aid for Young Scientists B (No. 15K17435) from JSPS.
\end{acknowledgments}

\appendix
%%%%%%%%%%%%%%%%%%%%%%%%%%%%%%%
\section{Details of the coupling functions}
\label{sec:app}
%%%%%%%%%%%%%%%%%%%%%%%%%%%%%%%
We first outline the derivation of Eq.~(\ref{eq:Cee_iso}) from Eq.~(\ref{eq:coupling_e-e}). Due to the spherical symmetry, it is reasonable to express the summation in Eq.~(\ref{eq:Cee_iso}) by
\begin{eqnarray}
 \sum_{\bm{k}} \rightarrow \frac{\Omega}{(2\pi)^3}
 \int_{0}^{\infty} dk k^2
 \int_{0}^{\pi} d \theta_{\bm{k}} \sin \theta_{\bm{k}} 
 \int_{0}^{2\pi} d\phi_{\bm{k}}.
\end{eqnarray}
Similarly, the summations with respect to $\bm{k}'$ and $\bm{q}$ are transformed into the integrals. Without loss of generality, we consider that $\bm{q}$ is parallel to the $z$-axis. Then, $\theta_{\bm{k}}$, $\phi_{\bm{k}}$, $\theta_{\bm{q}}$, and $\phi_{\bm{q}}$ can be regarded as the relative angle for $\bm{q}$. Then, one obtains
\begin{eqnarray}
& & C_{\rm e \mathchar`- e}(\varepsilon,\varepsilon'; \xi,\xi') 
\nonumber\\
 &=& \frac{4\pi^3}{\hbar {\cal N}(\varepsilon)}
 \left(\frac{2m^2}{\hbar^4}\right)^2
 \int_{0}^{\pi} d \theta_{\bm{k}} 
 \int_{0}^{\pi} d \theta_{\bm{k}'}
\int_{0}^{\infty} dq \vert \tilde{V}(\bm{q}) \vert^2
\nonumber\\
&\times&
\delta (\theta_{\bm{k}}  - \theta_{\bm{k}}^{0}) 
\delta (\theta_{\bm{k}'}  - \theta_{\bm{k}'}^{0}),
\end{eqnarray}
where $\theta_{\bm{k}}^{0}$ and $\theta_{\bm{k}'}^{0}$ are given by
\begin{eqnarray}
\theta_{\bm{k}}^{0} &=& 
\arccos\left[ \frac{\sqrt{2m}}{2\hbar q\sqrt{\varepsilon}}
(\xi - \varepsilon - \varepsilon_{\bm{q}})\right],
\nonumber\\
\theta_{\bm{k}'}^{0} &=& 
\arccos\left[ - \frac{\sqrt{2m}}{2\hbar q\sqrt{\varepsilon'}}
(\xi' - \varepsilon' - \varepsilon_{\bm{q}})\right],
\end{eqnarray}
respectively. If $0 \le \theta_{\bm{k}}^{0}\le \pi$ and $0 \le \theta_{\bm{k}'}^{0}\le \pi$, the integration with respect to $\theta_{\bm{k}}$ and $\theta_{\bm{k}'}$ becomes unity. Thus, one obtains Eq.~(\ref{eq:Cee_iso}) in the main text. It would be straightforward to derive Eq.~(\ref{eq:Cep_iso}) from Eq.~(\ref{eq:coupling_e-ph}). 

We next consider the numerical implementation of Eq.~(\ref{eq:coupling_ph-ph}). Using the spherical coordinates, we express the phonon wavevector and the polarization vectors as follows 
\begin{eqnarray}
 \bm{Q} &=& Q \bm{e}_r,
\nonumber\\
 \bm{e}(\bm{Q},{\rm LA}) &=& \bm{e}_r = (\sin\theta \cos\phi, \sin\theta \sin\phi, \cos\theta),
\nonumber\\
 \bm{e}(\bm{Q},{\rm TA1}) &=& \bm{e}_\theta = (\cos\theta \cos\phi, \cos\theta \sin\phi, - \sin\theta),
\nonumber \\
 \bm{e}(\bm{Q},{\rm TA2}) &=& \bm{e}_\phi = (- \sin \phi, \cos\phi, 0),
\end{eqnarray}
where the abbreviation of $\theta = \theta_{\bm{Q}}$ and $\phi = \phi_{\bm{Q}}$ is used. Similarly, we define 
\begin{eqnarray}
 \bm{Q}' &=& Q' \bm{e}'_r,
\nonumber\\
 \bm{e}(\bm{Q}',{\rm LA}) &=&  (\sin\theta' \cos\phi', \sin\theta' \sin\phi', \cos\theta'),
\nonumber\\
 \bm{e}(\bm{Q}',{\rm TA1}) &=&  (\cos\theta' \cos\phi', \cos\theta' \sin\phi', - \sin\theta'),
\nonumber \\
 \bm{e}(\bm{Q}',{\rm TA2}) &=& (- \sin \phi', \cos\phi', 0),
\end{eqnarray}
where $\theta' = \theta_{\bm{Q}'}$ and $\phi' = \phi_{\bm{Q}'}$. Using these expressions, we can obtain the expressions for $\bm{Q}+\bm{Q}'$ and $\bm{e}(\bm{Q} + \bm{Q}',\gamma'')$ in terms of $Q, Q', \theta, \theta', \phi$, and $\phi'$. Then one obtains 
\begin{eqnarray}
 & & 
 C_{\rm ph \mathchar`- ph}(\omega,\omega',\omega'',\gamma,\gamma',\gamma'') 
 \nonumber\\
 &=& \frac{\Omega}{(2\pi)^6\hbar \cal{D}_\gamma (\omega)} 
 \left(\frac{\hbar}{2\rho_{\rm i}}\right)^3 
 \int dS \int dS'
 \nonumber\\
&\times&
 \frac{\omega\omega'}{(v_\gamma v_{\gamma'})^3}
\frac{\vert A_{\rm ph} \vert^2}{v_{\gamma''} \vert \bm{Q} + \bm{Q}'\vert }
\delta (\omega'' - v_{\gamma''}\vert \bm{Q} + \bm{Q}'\vert),
\label{eq:C_ph_ph_app}
\end{eqnarray}
where $\int dS = \int d\theta \sin\theta \int d\phi$. The numerical integrals for $\theta$ and $\phi$ are efficiently performed by using the spherical design \cite{spdesign}. In the present work, 86 points on a sphere were used. The Dirac-delta function in Eq.~(\ref{eq:C_ph_ph_app}) is approximated to the Gaussian function with the broadening of $0.02 E_0$. To reduce the numerical errors of the total energy, we used the symmetry properties of $f (\omega,\omega',\omega'',\gamma,\gamma',\gamma'') =  {\cal D}_\gamma (\omega) C_{\rm ph \mathchar`- ph}(\omega,\omega',\omega'',\gamma,\gamma',\gamma'')$ with the replacement of $(\omega,\gamma) \leftrightarrow (\omega',\gamma')  \leftrightarrow (\omega'',\gamma'')$. Using the averaged value of $f$s, that is, 
\begin{eqnarray*}
& & 
\frac{1}{6}\Big[
f (\omega,\omega',\omega'',\gamma,\gamma',\gamma'')
+ f (\omega,\omega'',\omega',\gamma,\gamma'',\gamma')
\nonumber\\
&+& 
f (\omega',\omega,\omega'',\gamma',\gamma,\gamma'')
+ f (\omega',\omega'',\omega,\gamma',\gamma'',\gamma)
\nonumber\\
&+&
 f (\omega'',\omega,\omega',\gamma'',\gamma,\gamma')
+ f (\omega'',\omega',\omega,\gamma'',\gamma',\gamma)
\Big],
\end{eqnarray*}
we redefine the ph-ph coupling function as $C_{\rm ph \mathchar`- ph}(\omega,\omega',\omega'',\gamma,\gamma',\gamma'') = f (\omega,\omega',\omega'',\gamma,\gamma',\gamma'') / {\cal D}_\gamma (\omega)$.
%===================================================================%
%   References
%===================================================================%

\end{document}